\documentclass[aoas]{imsart}

\RequirePackage{amsthm,amsmath,amsfonts,amssymb}
\RequirePackage[authoryear]{natbib}

\usepackage[utf8]{inputenc} 
\usepackage[T1]{fontenc}    
\usepackage{hyperref}       
\usepackage{url}            
\usepackage{booktabs}       
\usepackage{amsfonts}       
\usepackage{nicefrac}       
\usepackage{microtype}      
\usepackage{xcolor}         
\usepackage{amsthm,amsmath,amssymb,mathtools}
\usepackage{latexsym}
\usepackage{graphics}
\usepackage{graphicx}
\usepackage{url}
\usepackage{hyperref}
\usepackage{bm}

\usepackage{subfigure}

\usepackage{caption}

\usepackage{enumitem}
\DeclareMathOperator*{\argmax}{arg\,max}
\DeclareMathOperator*{\argmin}{arg\,min}

\def\0{\boldsymbol 0}
\def\1{\boldsymbol 1}
\def\2{\boldsymbol 2}
\def\3{\boldsymbol 3}
\def\4{\boldsymbol 4}
\def\5{\boldsymbol 5}
\def\6{\boldsymbol 6}
\def\7{\boldsymbol 7}
\def\8{\boldsymbol 8}
\def\9{\boldsymbol 9}
\def\a{\boldsymbol a}

\def\f{\boldsymbol f}

\def\u{\boldsymbol u}

\def\x{\boldsymbol x}
\def\y{\boldsymbol y}
\def\z{\boldsymbol z}
\def\A{\boldsymbol A}
\def\B{\boldsymbol B}

\def\F{\boldsymbol F}

\def\I{\boldsymbol I}

\def\Q{\boldsymbol Q}
\def\R{\mathbb{ R}}

\def\U{\boldsymbol U}

\def\W{\boldsymbol W}
\def\X{\boldsymbol X}

\def\Z{\boldsymbol Z}

\def\|{\Vert}

\def\1{\mathbbm{1}}

\startlocaldefs

\endlocaldefs

\begin{document}

\begin{frontmatter}
\title{Feature Augmentations for High-Dimensional Learning: Applications to Stock Market Prediction Using Chinese News Data}
\runtitle{Feature Augmentations for High-Dimensional Learning}

\begin{aug}
\author[A]{\fnms{Xiaonan}~\snm{Zhu}\ead[label=e1]{xz8451@princeton.edu}},
\author[A]{\fnms{Bingyan}~\snm{Wang}\ead[label=e2]{bingyanw@princeton.edu}}
\and
\author[A]{\fnms{Jianqing}~\snm{Fan}\ead[label=e3]{jqfan@princeton.edu}}
\address[A]{Department of Operations Research and Financial Engineering, Princeton University\printead[presep={,\ }]{e1,e2,e3}}

\end{aug}

\begin{abstract}
High-dimensional measurements are often correlated which motivates their approximation by factor models. This holds also true when features are engineered via low-dimensional interactions or kernel tricks.  This often results in over parametrization and requires a fast dimensionality reduction.  We propose a simple technique to enhance the performance of supervised learning algorithms by augmenting features with factors extracted from design matrices and their transformations. This is implemented by using the factors and idiosyncratic residuals which significantly weaken the correlations between input variables and hence increase the interpretability of learning algorithms and numerical stability.
Extensive experiments on various algorithms and real-world data in diverse fields are carried out, among which we put special emphasis on the stock return prediction problem with Chinese financial news data due to the increasing interest in NLP problems in financial studies.
We verify the capability of the proposed feature augmentation approach to boost overall prediction performance with the same algorithm.
The approach bridges a gap in research that has been overlooked in previous studies, which focus either on collecting additional data or constructing more powerful algorithms, whereas our method lies in between these two directions using a simple PCA augmentation.

\end{abstract}

\begin{keyword}
\kwd{Factor Augmentations}
\kwd{Principal Components}
\kwd{Feature Interactions}
\kwd{Kernel Features}
\kwd{Prediction}
\end{keyword}

\end{frontmatter}


%


\section{Introduction}
Supervised learning has been an active area of research over decades, aiming to reveal underlying patterns in big data to enhance prediction accuracy. Previous studies primarily focus on two key areas: collecting extensive data from various sources and developing advanced algorithms to leverage the data.  
While substantial progress has been made in these areas, an intermediary aspect has received limited attention: the potential for enriching data features before feeding them into learning models. Note that various features can be extracted from the data and subsequently augmented to increase the prediction accuracy, provided that the augmented signals dominate the noises due to variable additions.
This approach is especially useful when a systematic method is available with minimal additional effort compared to the potential gain in model performance. It is highly versatile and can be applied independently of other methods or algorithms.

Considering the pair of $(\x,y)$, the response variable $y$ can be viewed as a part of the covariates $\x$ since $\x$ is collected to contribute to estimate $y$. Additionally, in practice, correlation effects often exist among observed features. Hence, there exist some latent common factors carrying the dependent structure of $\x$ and shared patterns between $\x$ and $y$. The most critical factors among $\x$ are also important to $y$. Meanwhile, since all the features are correlated with $y$, as the number of features increases to high-dimensional realms, the factors that influence $y$ become more significant.
Various approaches have been explored to estimate an approximate factor model (see, e.g., \cite{fama1992cross, stock2002macroeconomic, bai2002determining, bai2003inferential}), 
and factor models have proven beneficial in various contexts, enhancing variable selection and model performance both theoretically and empirically \citep{wang2019factor, stock2002macroeconomic}. From the methodological perspective, the Factor-Adjusted Regularized Model Selection (FarmSelect) approach proposed by \cite{fan2020factor} addresses sparse regression-related problems for model selection using latent factors to reduce variable dependence. Similarly, for prediction purposes, \cite{zhou2021measuring} proposes the Factor-Augmented Regularized Model for Prediction (FarmPredict) to analyze house usage using the integration of nightlight data and land planning data.  
Moving from linear to nonlinear prediction, \cite{fan2024factor} proposes the Factor Augmented Sparse Throughput (FAST) model that utilizes factor models for nonparametric variable selection and regression with deep ReLU neural networks.

To our knowledge, although being frequently used, factor models are so far only applied to the design matrix/tensor directly. No systematic studies have been made on the potential of information gain of such an approach that leads to systematic improvements in prediction power.
On the contrary, we propose to extract nonlinear factors from transformed versions of the design matrix, like low-dimensional interactions and kernel tricks. 
These transformations are likely to contain different patterns and valuable information compared to the original data.
Augmenting features with factors derived from these transformations elevates the feature space to higher dimensions, enhancing approximation power. This approach not only produces a better interpretation of the data but also aids variable selection in high-dimensional cases by decreasing feature correlations.
Meanwhile, since only several more features are added (say, 5 to 10) via principal component analysis, the variance they bring about is almost negligible compared to the number of features in the problems. These two aspects together lead to the benefits of the proposed method in statistical prediction, as to be demonstrated.

Building on this methodological foundation, we investigate its performance in a real-world financial prediction setting involving Chinese news text data. Natural Language Processing (NLP) is gaining increasing prominence nowadays in diverse applications.
Recent studies have highlighted the value of textual information in financial modeling and decision-making \citep{fan2021much, ke2019predicting, goldstein2021big, wang2024application, loughran2016textual}. These works primarily concentrate on the collection and preprocessing of informative text data, the transformation of textual content into structured, machine-readable inputs, and the development of predictive models to address tasks such as asset pricing and risk assessment.
In many cases, especially in financial studies, the data is tedious to collect and expensive to buy. Therefore, it is very useful and helpful to be able to extract some latent and informative features, preferably in an easy way that does not require much additional computation, and achieve better estimation performance.

To present that, we leverage the large-scale Chinese financial news dataset compiled by \cite{fan2021much} and study the stock market problems.
The original paper made great efforts on collecting and processing the news data, after which the authors applied the Factor-Adjusted Regularized Model Selection \citep{fan2020factor} with linear regression on the stock return, and analyzed thoroughly, focusing on the sentiment scores and portfolio returns, from a finance perspective. 
In contrast, our focus lies in demonstrating the simple and general statistical augmentation technique of extracting latent non-linear factors from the high-dimensional embedded textual data, and showing how such augmentations can enhance stock return prediction in financial market studies. We apply this approach across five widely used machine learning algorithms—Lasso, Ridge Regression, Random Forests, Gradient Boosted Trees, and Neural Networks—and consistently observe improved performance, underscoring the robustness and flexibility of the method. Building on these estimation results, we further conduct an event study and a portfolio analysis to illustrate the practical value of the proposed feature augmentation framework in the context of financial investment.

Furthermore, to highlight the versatility of the proposed method, we supplement our primary case study with additional empirical evaluations spanning a variety of domains and problem. These supplementary experiments confirm that the augmentation framework consistently enhances predictive accuracy in diverse applications, reinforcing its general applicability in high-dimensional learning contexts. In addition, the results offer practical guidance for selecting appropriate transformation methods tailored to different problem settings.


The remainder of the paper is organized as follows. Section~\ref{sec:data} introduces the dataset of Chinese financial news and the associated stock returns. Section~\ref{sec_methodology} outlines the proposed methodology, including matrix transformations, factor estimation, feature augmentation, variable screening, and the learning algorithms. Section \ref{sec:chinese_result0} presents the empirical analysis of the Chinese News data, and as a supplement, Section \ref{sec_empirical} provides additional empirical studies across diverse datasets, illustrating the versatility and effectiveness of the method. Finally, Section~\ref{sec_conclusion} concludes the paper and discusses potential directions for future research.

Some notations used are as follows. Bold uppercase letters, bold lowercase letters, and unbold lowercase letters represent matrices, vectors, and scalars, respectively. Letters with hats are estimators. $\I_n\in\mathbb R^{n\times n}$ is the identity matrix, and $\mathbf{1}_n \in \mathbb{R}^n$ represents the all-one vector.


\section{Data Description and Pre-processing}\label{sec:data}

We utilize the large-scale Chinese financial news dataset constructed by \citet{fan2021much}, collected from Sina Finance, one of the leading financial news platforms in China. Detailed information about the dataset can be found therein; we provide a brief summary and pre-processing procedure below.

The dataset spans from 2000 to 2019 and includes at most 300 articles per day, resulting in a final corpus of approximately 914,000 articles. Each news article was crawled along with its associated publication time and corresponding stock information, and paired with the effective beta-adjusted return of the linked stock on the day of publication, which serves as the target response variable in our study. More specifically, the matching stock is decided based on a combination of html and article content, and the time range of daily return is set as the close-to-close return covering the article’s publish time. 
The title and main text of each article are segmented into lists of words and phrases using the Jieba Python package \citep{sun2017jieba}, resulting in a bag of words representation for each article and collecting a vocabulary of approximately 1,181,000 unique words and phrases. Therefore, for each article, a vector of the same length ($1,181,000$) is generated, with each element being the word count of the corresponding word that appears in the article.

We adopt a six-month-ahead rolling window estimation strategy, whereby for each window, ten years of data are used for model training, followed by a six-month testing period. Given the high dimensionality and sparsity of the text data, we implement a two-step dimension reduction procedure in each training window. First, we retain only the 3,000 most frequent words in the training set. Then, we apply a screening technique, as detailed in Section~\ref{sec:screen}, to further reduce the feature set to a comprehensive subset of 300 words and phrases for subsequent modeling.

\section{Methodology}\label{sec_methodology}
In this section, we present the general framework of feature augmentation with nonlinear latent factors and specify the application on the news text dataset.

\subsection{Data Transformation} \label{sec_mx_transform}
One often applies nonlinear transforms to create additional features and
results in overparametrization and requires some dimensionality reduction.
The factor model extracts information from a large matrix and creates low-dimensional features.  Unlike prior literature that applies the factor model directly to the data matrix, we extend it to the simple transformations of the data. Since the process of nonlinear transformations may reveal additional nonlinear features, it is expected that extracted factors can capture distinct aspects of the data compared to those obtained directly from the original data.
There are many kinds of nonlinear data transformations. We showcase three methods: interaction matrices, kernel matrices, and the intermediate outputs of neural networks. 

\vspace{0.5em}
\noindent\textbf{Pairwise interactions: } Let $\X = (\x_1,\ldots,\x_n)^\top$ be the original feature matrix with $\x_{\! i}\!\! =\! (\!x_{i1},\!\ldots\!, x_{ip})^{\!\top}$. Then we augment the original feature matrix $\X$ with $\X_{\mathrm{inter}}\in\mathbb{R}^{n\times p(p+1)/2}$ which is defined as
$$\X_{\mathrm{inter}} := (\mathsf{vec}(\{x_{1i}x_{1j}\}_{1\leq i< j\leq p}),\ldots,\mathsf{vec}(\{x_{ni}x_{nj}\}_{1\leq i< j\leq p}))^\top,$$
where $\mathsf{vec}(\cdot)$ is the vectorization operator. 
This transformation reveals latent semantic relationships between words. For example, while the individual counts of "artificial" and "intelligence" may not be highly informative alone, their co-occurrence can strongly indicate a specific topic, and factors extracted therein enrich the feature space to capture concept combinations that are not visible through single-word frequencies alone. 
Interaction terms are also useful in many other problems, such as the study of gene synergy, protein interactions, image pixels interplay, or word meanings, among others \citep{wu2016regularized, balli2013interaction, zhang2021building}. Thus, it is reasonable to expect that factors derived from the interaction matrices are able to provide additional information, enhancing the capabilities of machine learning models.

\vspace{0.5em}
\noindent\textbf{Kernel methods:}  Kernel methods allow us to examine higher order interactions through a similarity kernel.  Gaussian and polynomial kernels are considered herein, whose similarity matrix are denoted respectively by $\X_{\mathrm{rbf}}$ and $\X_{\mathrm{poly}}$. Since the number of samples $n$ is extremely large in each training window (approximately 300{,}000 in our main study), we randomly select $n_0$ columns, resulting in a reduced kernel matrix of size $n \times n_0$, to learn latent factors.  This significantly reduces the computational burden while preserving the essential geometry of the feature space induced by the kernel.

By implicitly projecting the original feature vectors into richer, nonlinear spaces, these methods allow us to uncover complex relationships that are not visible in the raw data. 
In the context of Chinese news text data, kernel-based transformations can capture nuanced semantic patterns—for example, articles using different expressions like “policy easing” and “interest rate cuts” may be recognized as discussing similar monetary events. Such nonlinear representations are capable of identifying topic clusters that transcend surface word overlap, capturing sentiment shifts expressed through unusual or indirect phrasing, and modeling higher-order interactions between terms that indicate subtle narrative themes. These latent semantic structures enhance the model’s ability to generalize and improve predictive accuracy beyond what is achievable through linear representations alone.

\vspace{0.5em}
\noindent\textbf{Neural networks:}
As for neural networks, we draw the last hidden layer of an FNN  applied to $\X$ and obtain ${\X}_{\mathrm{fnn}}$. Shallow neural networks with one or two layers of convolution with ReLU activation with a wide enough last hidden layer output is enough to be combined by PCA to derive the factors.
In the context of news text data for stock prediction, neural networks can capture complex interactions between terms—such as "interest rates," "central bank," and "market volatility"—that individually provide limited information but together reveal important market signals. By modeling such nonlinear word combinations, neural networks help generate richer feature representations that better align with the underlying drivers of stock returns.
Moreover, as the method is widely applicable, when considering image data, we may also use convolutional neural networks (CNN), and extract ${\X}_{\mathrm{cnn}}$ from its last hidden layer.
Considering the widespread empirical success of neural networks in uncovering intricate and nonlinear data structures, this transformation is expected to produce powerful and informative features that complement the original representation.

\subsection{Factor Model}
Suppose that there are $n$ samples $\{\z_i\}^n_{i=1}$, $\z_i\in\mathbb{R}^{p}$ generated from the factor model
\begin{equation}\label{factormodel1}
{\z}_{i} =
\a+\B\f_i+{\u}_i,
\end{equation}
where $\B\in\mathbb{R}^{p\times K}$ 
is a factor loading matrix, and the latent factors $\f_i\in\mathbb{R}^{K}$ are zero-mean random variables, uncorrelated with the idiosyncratic components ${\u}_i \in \mathbb{R}^{p}$.
The intercept $\a$ is estimated by the average of samples, i.e.  $\widehat{\a}=\overline{\z}= \sum_{i=1}^n \z_i/n$. Without loss of generality, we assume that $\widehat{\a}=\mathbf{0}$.
Denote the design matrix by $\Z=(\z_1,\ldots,\z_n)^\top\in\mathbb R^{n\times p}$.

\subsubsection{Factor Estimation via PCA}\label{Sec_erbd}
A standard way of extracting factors from a matrix is to apply PCA to the whole data matrix.
For identifiability purposes, 
we assume that $\operatorname{cov}(\f_i)=\I_K$ and ${\B}^{\top}{\B}$ is diagonal. 
By applying PCA on the sample covariance of $\left\{\z_i\right\}_{i=1}^n$, the factors can be estimated by \citep{bai2003inferential}
$$
\widehat{\B}=\big({\widehat\lambda_1}^{1/2}\widehat{\bm\xi}_1\ldots,{\widehat\lambda_K}^{1/2}\widehat{\bm\xi}_K\big),
\text{ and }
\widehat{\f}_i = \mathrm{diag}\big(\widehat\lambda_1,\ldots,\widehat\lambda_K\big)^{-1}\widehat{\B}^\top\z_i,
$$
where $\widehat\lambda_1,\ldots,\widehat\lambda_K$ and $\widehat{\bm\xi}_1,\ldots,\widehat{\bm\xi}_K$ are the top $K$ eigenvalues of the sample covariance matrix and their associated eigenvectors.
Let $\widehat{\F}:=(\widehat{\f}_1,\ldots,\widehat{\f}_n)^\top$ be the matrix of estimated latent factors. 
This is the same as setting the columns of $\widehat{\F}/\sqrt{n}$ as the eigenvectors of $\Z\Z^\top$ corresponding to the top $K$ eigenvalues and letting $\widehat{\B}=n^{-1}\Z^\top\widehat{\F}$.

There are several methods to determine $K$, the number of factors, 
such as the adjusted eigenvalues thresholding method \citep{fan2022estimating} and the eigenvalue ratio (eigen-ratio) estimator \citep{tsai2009ebv, lam2012factor,ahn2013eigenvalue}, to list just a few.
Note that our down stream prediction task is not very sensitive to the slight over estimation of  $K$.
Here, we adopt the eigen-ratio method with $k_{\max}=\lfloor (p\wedge n)/3 \rfloor$ suggested by \cite{ahn2013eigenvalue} and $k_{\min}=\operatorname{max}\{\lfloor (p\wedge n)/10 \rfloor,2\}$:
\[
\widehat K := \argmax_{k_{\min}\leq j \leq k_{\max}} {\widehat{\lambda}_j}/{\widehat{\lambda}_{j+1}}.
\]
Note that the lower and upper bound is heuristic, and it may not be essential in some cases where some conditions are met \citep{zhang2022non}. Also, if $p\wedge n$ is large, we can set $k_{\min}$ to, for example, $5$, just to avoid degenerate outcomes \citep{wainwright2019high, marchenko1967distribution}.

\subsubsection{Factor Estimation via Diversified Projection}\label{Sec_dp}
Though powerful and popular over the years, the principal component estimation of factors has potential drawbacks.  One big issue is that it is computationally expensive for high-dimensional features and also requires a large sample size to get accurate estimates. 
This computational scalability issue becomes very severe in text data, where the dimension of the input matrix can be very large. 
To sidestep this problem, \cite{fan2022learning} propose to learn latent factors using diversified projections by weighted averages with predetermined weights. Later on, \cite{fan2024factor} propose a data-driven pre-trained weight matrix, calculated by applying PCA to a separate set of $n^\prime$ examples.  It is formally derived there that $n^\prime$ can be much smaller than the original sample size $n$: it suffices to have $n^\prime \asymp K^2 \log p$.
More specifically, let $\{\z^\prime_i\}_{i=1}^{n^\prime} \subseteq \mathbb{R}^p$ be another $n^\prime$ samples that are independent with $\{\z_i\}_{i=1}^{n}$. Then PCA is applied to get the top-$K^\prime$ eigenvectors $\widehat{\bm\xi}^\prime_1,\ldots,\widehat{\bm\xi}^\prime_{K^\prime}$ of the sample covariance matrix of $\{\z^\prime_i\}_{i=1}^{n^\prime}$, where $K^\prime\geq K$. Choose the diversified weight matrix as $\W\!=\!\sqrt{p}\, (\widehat{\bm\xi}^\prime_1,\ldots,\widehat{\bm\xi}^\prime_{K^\prime})$.
With the pre-training in hand, we can proceed as if the factors are known to us, and the loading matrix $\B$ can thereafter be estimated by least squares:
\[
\widehat{\f}_i :=  \W^\top\z_i/p, \text{ and }
\widehat{\B} = \sum_{i=1}^n\z_i \widehat{\f}_i^\top\big(\sum_{i=1}^n\widehat{\f}_i\widehat{\f}_i^\top\big)^{-1}.
\]

Besides the aforementioned advantages, diversified projection offers two additional benefits for real-world applications, especially in problems like the studied textual data example, where both the input dimension and sample sizes are very large. First, when the dimension $p$ is large, using a small subset of data with a reduced sample size $n'$ can significantly accelerate the estimation of the number of factors, since the nonzero eigenvalues of $\Z \Z^\top$ and $\Z^\top \Z$ are identical, and their corresponding eigenvectors are related through simple linear transformations. Second, when the sample size $n$ (or $n_0$) is ultra-large, applying PCA to the full kernel matrix becomes computationally infeasible, whereas diversified projection remains computationally tractable and effective.

\subsection{Feature augmentation} \label{Sec_feature_aug}
Denote by $\widehat{\F}_0$, $\widehat{\F}_\mathsf{inter}$, $\widehat{\F}_\mathsf{poly}$, $\widehat{\F}_\mathsf{rbf}$, and  $\widehat{\F}_\mathsf{fnn}$
the latent factors extracted respectively from $\X$ and the transformed matrices $\X_\mathsf{inter}$, $\X_\mathsf{poly}$, $\X_\mathsf{rbf}$, and $\X_\mathsf{fnn}$
introduced in Section \ref{sec_mx_transform}. We demean $\X$ and all the transformed matrices before further processing. In what follows, we take $\widehat{\F}_\mathsf{inter}$ as an example, while the method can also be applied to other factors. 

We augment the original feature space $\X$ by adding $\widehat\F_\mathsf{inter}$. To decorrelate $\X$ from $\widehat\F_\mathsf{inter}$ while retaining all the information, we project $\X$ onto the subspace spanned by the factors $\widehat\F_\mathsf{inter}$, i.e. fit the model
\begin{align}
\label{factormodel2}
\X &= \widehat\F_\mathsf{inter}\B_\mathsf{inter}^\top + \U_\mathsf{inter},
\end{align}
and obtain the estimates
 \begin{align}
\widehat{\B}_\mathsf{inter} &= \left[(\widehat\F_\mathsf{inter}^\top\widehat\F_\mathsf{inter})^{-1}\widehat\F_\mathsf{inter}^\top\X\right]^\top, \\
\widehat{\U}_\mathsf{inter}&=[\I_n-\widehat\F_\mathsf{inter}(\widehat\F_\mathsf{inter}^\top\widehat\F_\mathsf{inter})^{-1}\widehat\F_\mathsf{inter}^\top]\X.
\end{align}
The subspace spanned by the new independent variables 
$(\widehat{\F}_\mathsf{inter}, \widehat{\U}_\mathsf{inter})\in\mathbb{R}^{n\times(K+p)}$ is the same as that of $(\widehat{\F}_\mathsf{inter}, \X)$. Note that when the factors are estimated by PCA, we have $\widehat\F_\mathsf{inter}^\top\widehat\F_\mathsf{inter}=\I$, and therefore $\widehat{\U}_\mathsf{inter}=[\I_n-\widehat\F_\mathsf{inter}\widehat\F_\mathsf{inter}^\top]\X$.

Let $y_i$ be the 
variable to be predicted based on $\x_i$. 
Then, the general regression model takes the form
\begin{align}
y_i = g(\widehat{\f}_{\mathrm{inter},i},\widehat{\u}_{\mathrm{inter},i}) + \varepsilon_i, \label{eq3}
\end{align}
where $\varepsilon_i$ is the idiosyncratic noise. Any statistical machine learning model can be employed here, such as Lasso Regression, Ridge Regression, Random Forest, Gradient Boosted Tree, and Neural Networks.
Moreover, different factors may be combined together to boost performance. Specifically, one can add $\widehat{\F}_0$ to $(\widehat{\F}_\mathsf{inter},\widehat{\U}_\mathsf{inter})$ for further augmentation, which gives
\[
y_i = \widetilde g (\widehat{\f}_{0,i}, \widehat{\f}_{\mathrm{inter},i},\widehat{\u}_{\mathrm{inter},i}) + \varepsilon_i.
\]
Note that we expand the original features to use principal component (factor) directions.  This significantly reduces possible modeling biases while not dramatically increasing the number of variables. 

Under the proposed model, the prediction for a given new feature vector $\x_{\mathrm{new}}$ consists of two steps. First, we compute the interactions for $\x_{\mathrm{new}}$ and denote it by $\x_{\mathrm{new}\!,\mathsf{inter}}$.
With the estimated factor loading matrix $\widehat{\B}_\mathsf{inter}$, estimating the latent factors corresponding to $\x_{\mathrm{new}}$ can be equivalently viewed as regressing $\x_{\mathrm{new},\mathsf{inter}}$ on $\widehat{\B}_\mathsf{inter}$ based on \eqref{factormodel1}, which gives
\[
\begin{aligned}
\widehat\f_{\mathrm{new}}
&=\big(\widehat{\B}_\mathsf{inter}^{\top}\widehat{\B}_\mathsf{inter}\big)^{-1}\widehat{\B}_\mathsf{inter}^{\top}\x_{\mathrm{new},\mathsf{inter}}\\
&=\operatorname{diag}\big(\widehat\lambda_1,\ldots,\widehat\lambda_K\big)^{-1} \widehat{\B}_\mathsf{inter}^{\top}\x_{\mathrm{new},\mathsf{inter}}.
\end{aligned}
\]
Alternatively, if the factors are estimated by diversified projection, we have 
\[
\widehat{\f}_{\mathrm{new}} := \frac{1}{p} \W^\top\x_{\mathrm{new},\mathsf{inter}}.
\]
Thus, by \eqref{factormodel2}, the new idiosyncratic component is
$\widehat\u_{\mathrm{new}}=\x_{\mathrm{new}}-\widehat\B_\mathsf{inter}\widehat\f_{\mathrm{new}}$, and
accordingly, $\widehat y_{\mathrm{new}}=\widehat g(\widehat\f_{\mathrm{new}},\widehat\u_{\mathrm{new}})$, where $\widehat g$ is the fitted model. 

\subsection{Decorrelated Variable Screening} \label{sec:screen}
To improve computational efficiency, conditional correlation screening can be applied to $\widehat{\u}_\mathsf{inter}$ to expeditiously select its useful components if the dimension $p$ is ultrahigh.
The screening process involves reducing the dimension by only selecting the features that are the most strongly correlated with the response variable. This technique was first proposed by \cite{fan2008sure} and later improved by \cite{fan2020factor}.
With the latent factors given, features of the residual are ranked according to their conditional marginal contributions.
Let $L_n(\y,\widehat\y)$ be an empirical convex  loss function under consideration for model \eqref{eq3}.
The augmented marginal regression considers the following low-dimensional fit: 
\[
\widehat\theta_j = \argmin_{\bm\gamma\in\mathbb{R}^{\hat K},\theta\in\mathbb{R}} L_n\big(\y, \widehat\F_\mathsf{inter}\bm\gamma+\widehat{\u}_j\theta),
\]
where $\widehat\F_\mathsf{inter}$ can be replaced by any latent factors introduced before, 
and $\widehat{\u}_j$ is the $j$-th column of $\widehat{\U}$ that is standardized.
Then $\big\{\widehat\theta_j\big\}_{j=1}^p$ are sorted in their absolute values, and the screening deletes the variables corresponding to the smallest marginal marginal contributions.  The selected $\widehat{\u}_j$ and the factors are then fed into a statistical machine learning algorithm to train a model. 
This procedure can also be applied to the original $\X$, which learns the common factors of $\X$ and important $\widehat{\u}_j$ for prediction.

\subsection{Statistical Machine Learning Methods}
Five statistical machining learn methods are considered as examples to verify if factor augmentation can improve the accuracy of learning algorithms. 

\vspace{0.5em}
\noindent\textbf{Lasso Regression:} Lasso is a popular regression analysis method for its convexity and ability to produce sparse estimators. It imposes $\ell_1$-norm regularization on linear regression.
Take the pairwise interaction matrix for example, let $\widehat{\Q}_\mathsf{inter}:=(\mathbf{1}_n, \widehat\F_\mathsf{inter}, \widehat\U_\mathsf{inter})\in\mathbb{R}^{n\times (p+K+1)}$ and $\bm\theta \in\mathbb{R}^{p+K+1}$. Lasso imposes $\ell_1$-norm regularization on linear regression and the model yields
\[\widehat{\bm\theta}\in\argmin_{\bm\theta\in\mathbb{R}^{p+K+1}}\left\{\frac{1}{n}\|\y-\widehat{\Q}_\mathsf{inter}\bm\theta\|^2+\gamma_1 \|\bm\theta\|_1 \right\},\]
where $\y=\left(y_1,\ldots,y_n\right)^\top$, and $\gamma_1$ is the tuning parameter.

\vspace{0.5em}
\noindent\textbf{Ridge Regression:}
 Similar to Lasso, Ridge adds $\ell_2$-norm penalization on linear regression.  However, Ridge does not produce sparsity. Instead, it deals with collinearity issues via
\[\widehat{\bm\theta}\in\argmin_{\bm\theta\in\mathbb{R}^{p+K+1}}\left\{\frac{1}{n}\|\y-\widehat{\Q}_\mathsf{inter}\bm\theta\|^2+\gamma_2 {\|\bm\theta\|_2^2} \right\},\]
where $\gamma_2$ is the tuning parameter. 
It admits the explicit form
\[
\widehat{\bm\theta} = \big(\widehat{\Q}_\mathsf{inter}^\top\widehat{\Q}_\mathsf{inter}+\gamma_2\I_{p+K+1}\big)^{-1}\widehat{\Q}_\mathsf{inter}^\top\y.
\]

\vspace{0.5em}
\noindent\textbf{Random Forest (RF):}
Random forest is an ensemble learning method used for both classification and regression problems \citep{breiman2001random}. It involves aggregating the results of multiple independent decision trees, each of which uses tree representation to solve problems. Random forest uses bootstrap to generate the collection of decision trees and uses a randomly selected subset of predictors at each split to construct each tree.
The outcome of RF is the majority vote of all the trees for classification and average of all tree predictions.  

\vspace{0.5em}
\noindent\textbf{Gradient Boosted Trees (GBT):}
Gradient Boosted Trees are also an ensembling method that iteratively performs regression or classification by fitting classification and regression trees to the pseudo residuals of the previous fit and aggregate the outcomes of the prediciton.  We omit the detail.


\vspace{0.5em}
\noindent\textbf{Neural Networks:}
Neural networks have achieved tremendous success in recent years thanks to the availability of big data. Fully-connected Neural Networks (FNN) is one of the most basic learning models.
For given layer widths $n_0,\ldots,n_{L+1}$ and a simple nonlinear function $\sigma$, the architecture of FNN with depth $L$ takes the form
$$
\mathbf{h}^{(L+1)}=\mathbf{A}^{(L+1)} \circ \sigma \circ \mathbf{A}^{(L)} \circ \ldots \circ \sigma \circ \mathbf{A}^{(1)}(\x),
$$
where $\circ$ denotes the composition of two functions, and $\mathbf{A}^{\ell}:\mathbb{R}^{n_{\ell-1}}\rightarrow\mathbb{R}^{n_{\ell}}$ is an affine function defined by
\[
\A^{(\ell)}(\x) = \W^{(\ell)}\x +\mathbf{b}^{(\ell)}, \text{ for } i=1,\ldots,L,
\]
with $\mathbf{W}^{(\ell)}$ and $\mathbf{b}^{(\ell)}$ being the weight matrix and the intercept, respectively, associated with the $\ell$-th layer. 
Common choices for the activation function $\sigma$ are $\operatorname{ReLU}(t)=\max\{0,\, t\}$ and $\operatorname{tanh}(t)$ among others.
Besides, to prevent overfitting, \cite{hinton2012improving} proposed to randomly drop out subsets of hidden units.
We only consider shallow FNN for classifications, for example, two layers of ReLU activation followed by Dropout layers, respectively.

\section{Empirical Analysis on Chinese Text Data} \label{sec:chinese_result0}
We apply the aforementioned feature augmentation methods to the Chinese news data to predict stock returns and showcase how we can benefit from the proposed nonlinear factors in diverse machine learning algorithms.
All experiments are done on a 4-rack Intel compute cluster with NVIDIA A100 GPUs.

\subsection{Number of Factors}
\begin{figure*}[!htb]
	\begin{center}
		\captionsetup{width=\linewidth}
		\centerline{
			\subfigure{\includegraphics[width=0.418\textwidth]{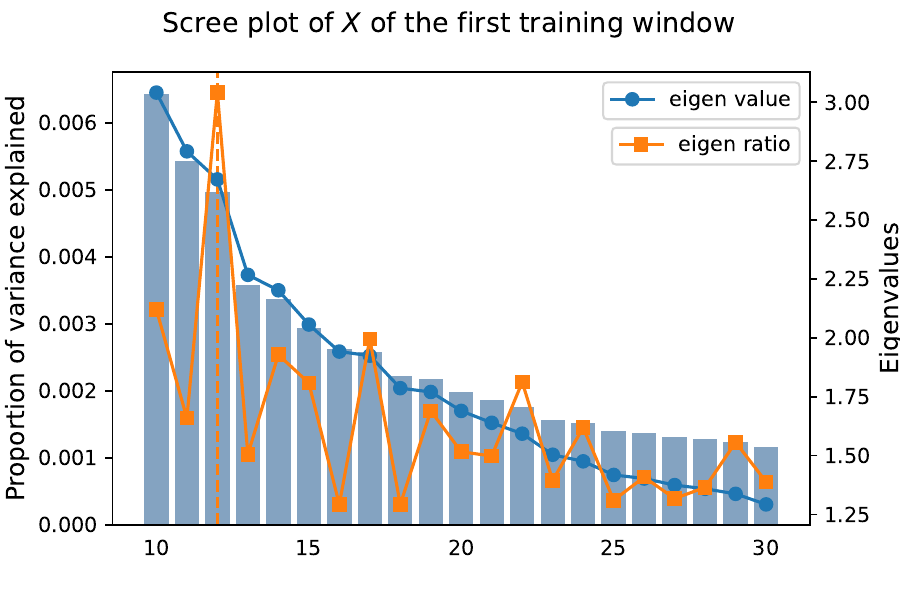}}
			\hspace{0.3cm}
			\subfigure{\includegraphics[width=0.41\textwidth]{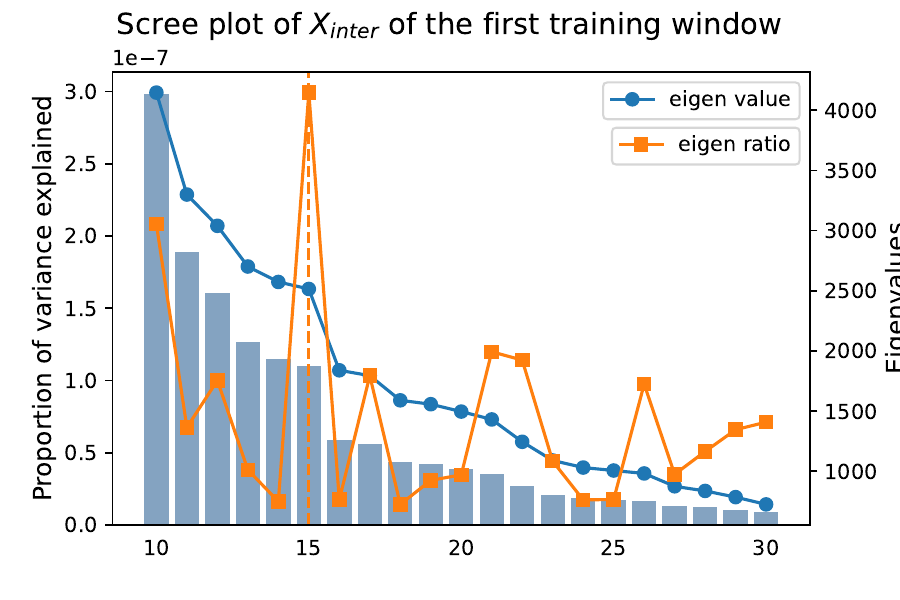}}
		}
		\setlength{\belowcaptionskip}{-20pt}
		\setlength{\abovecaptionskip}{0pt}
		\caption{Scree plot for covariance matrices of the original $\X$ (left) and the interaction matrix $\X_{\mathrm{inter}}$ (right) of the first training window of the Chinese News data. Eigenvalues (round dotted blue line), eigen-ratios (square dotted orange line), and proportion of variance explained (bar) by the top 10 to 30 eigenvectors}
		\label{Fig_K}
	\end{center}
\end{figure*}

Before presenting the results, we discuss the way of determining the number of factors $K$ for each transformation of the Chinese text data. We use the diversified projection method, illustrated in Section \ref{Sec_dp}, and set the pre-training set size to be $n' = 1000$ for each training window. We estimate $K$ based on the pertaining set data using the eigen-ratio method illustrated at the end sof Section \ref{Sec_erbd}.
Figure \ref{Fig_K} displays the scree plots of the top 10 to 30 eigenvalues, eigenvectors, and eigen-ratios of the covariance of $\X$ (the original matrix) and $\X_{\mathrm{inter}}$ (the interaction matrix), of the first training window. By the eigen-ratio method, the numbers of factors are chosen as 12 and 15, respectively, for $\F_0$ and $\F_{\mathrm{inter}}$. The same procedure is performed for all windows and models. With the number of factors determined, we present the results and findings as follows.

\subsection{Tuning and Testing}
Five machine learning techniques are employed for the classification problems, including Lasso, Ridge, RF, GBT, and FNN.
The hyperparameters for all the algorithms and factor models are selected by cross-validation (CV) with predefined grids. To be more specific, the maximum number of iterations 
to converge in Lasso and Ridge, the number of trees in RF, and the number of boosting rounds in GBT are all set to 100. 
The tuning parameter $\gamma_1$ for Lasso is chosen from $15$ numbers from $10^{-10}$ to $10^{-3}$ with even space on a log scale, and $\gamma_2$ for Ridge is chosen as a geometric sequence from $10$ numbers from $10^{-3}$ to $10^3$.
In RF, the maximum depth of trees and the minimum sample number required to split an internal node are both chosen from $(4,8,16)$.  In GBT, the maximum depth of trees is chosen from the same set, and the learning rate is chosen from $(0.2, 0.02)$.

The performance is evaluated using the out-of-sample $R^2$.
Let $(\x_t,y_t)$, $t=1,\ldots,n_{\mathrm{total}}$ be the time series data, where $n_{\mathrm{total}}$ is the total number of time points. Note that $\x_t$ are features known before time $t$.
Consider a rolling window prediction with a window size $m$. Each window consists of a consecutive observation of $m$ data points $(\x_{t-m},y_{t-m}),\ldots,(\x_{t-1},y_{t-1})$, and they are taken as the training set to predict $y_t$ based on $\x_t$.  To save the computation, the model is updated only once every $h$ predictions so that $y_t, \ldots,y_{t+h-1}$ are predicted based on $\x_t,\ldots,\x_{t+h-1}$, using the same trained model. After each group of predictions, the training window is shifted forward by $h$ time points to retrain the model, and the next $h$ out-of-sample time points are predicted.
Let $\widehat{y}_t$, $t=m+1,\ldots,n_{\mathrm{total}}$ be the prediction of rolling windows. Let $\overline{y}_t$ be the sample mean of the training set corresponding to the data point $t$. For example, $\overline{y}_{m+i} = \sum_{j=1}^{m}y_j/m$ for $i=1,\ldots,h$. 
Then, we can define the out-of-sample $R^2$ in the same way as above, which is
\[
R^2 = 1-\frac{\sum_{t=m+1}^{n_{\mathrm{total}}}\big(y_t-\widehat y_t\big)^2}{\sum_{t=m+1}^{n_{\mathrm{total}}}\big(y_t-\overline{y}_t\big)^2}.
\]

\subsection{Prediction Results}\label{sec:chinese_result}
We apply the proposed feature augmentation methods to the Chinese news text data for the purpose of predicting the returns of associated stocks. The primary objective is to assess the extent to which stock return prediction can benefit from the incorporation of augmented features. To ensure a fair comparison, all experimental settings are held constant except for one aspect: whether or not factor-based augmentations are included. Accordingly, we establish benchmark models that are applied directly to the original feature matrix $\X$ and compare their performance against augmentations with the proposed factors.
We consider five types of augmentation factors—$\widehat{\F}_\mathsf{inter}$, $\widehat{\F}_\mathsf{poly}$, $\widehat{\F}_\mathsf{rbf}$, and $\widehat{\F}_\mathsf{fnn}$—as introduced in Section~\ref{sec_mx_transform}. 
Hereafter, when we talk about $(\widehat{\F},\widehat{\U})$ for a factor $\widehat{\F}$, the matrix $\widehat{\U}$ is always the residual of $\X$ on $\widehat{\F}$, and we will not specify the corresponding factor type for simplicity.

\begin{figure*}[htp]
	\centering     
	\captionsetup{width=\linewidth}
	\includegraphics[width=0.85\textwidth]{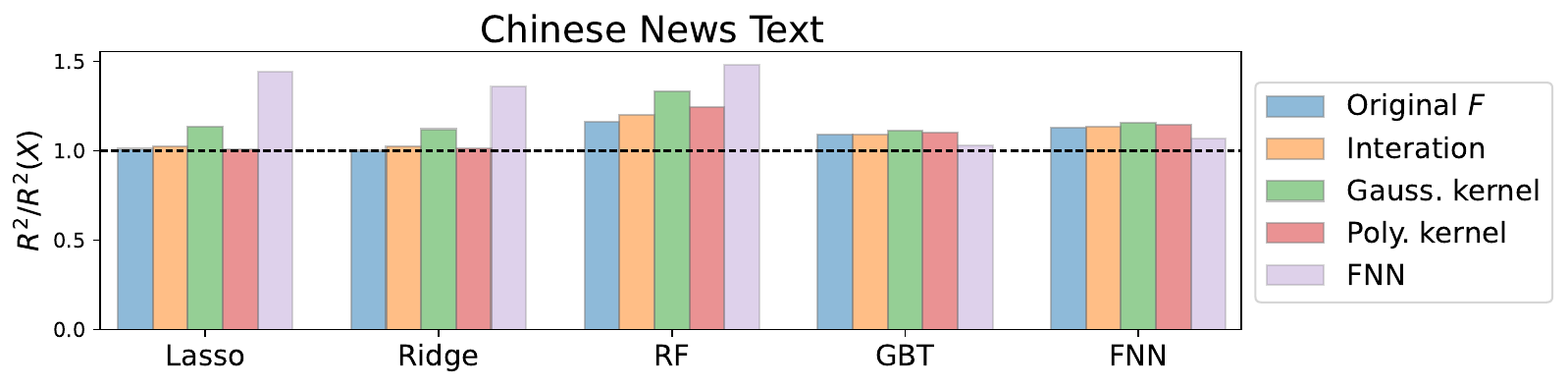}
	\caption{Ratio of the out-of-sample $R^2$ of each model to that without feature augmentation ($R^2(\X)$, benchmark) for Chinese news text dataset by diversified projection. The bars above the horizontal line at 1 indicate that the corresponding factor augmentation methods perform better than the benchmark.}\label{Fig_Chinese_new}
\end{figure*}

All experiments are repeated 20 times to reduce the influence of randomness in selecting the subsample of size 1000 to create different diversified projections, and the average results are reported.
Figure \ref{Fig_Chinese_new} demonstrates the relative performance of the feature augmentation methods. All the out-of-sample $R^2$ are divided by $R^2(\X)$, the out-of-sample $R^2$ of the benchmark model. Greater values indicate better performance.
The blue bars are for $(\widehat{\F}_0,\widehat{\U})$. For each of the newly proposed factors $\widehat{\F}$,
we consider adding factors $\widehat{\F}_0$ to further augment the feature space $(\widehat{\F},\widehat{\U})$. Two ways of aggregation are carried out. One is to directly use  $(\widehat{\F}_0,\widehat{\F},\widehat{\U})$ for prediction, and the other is to decorrelate $\widehat{\F}_0$ from $\widehat{\U}$ by consider $(\widehat{\F}_0,\widehat{\F},\widetilde{\U})$, where $\widetilde\U:=[\I_n-\widehat\F_0(\widehat\F_0^\top\widehat\F_0)^{-1}\widehat\F^\top]\widehat{\U}.$ These two methods are equivalent under linear regression but may yield different results under other models.
In Figure \ref{Fig_Chinese_new}, we present the best outcomes among the three augmentations $(\widehat{\F},\widehat{\U})$, $(\widehat{\F}_0,\widehat{\F},\widehat{\U})$, and $(\widehat{\F}_0,\widehat{\F},\widetilde{\U})$.

The results consistently show that feature augmentation with latent factors enhances predictive performance relative to the benchmark. In many cases, nonlinear factors perform comparably to or significantly better than the linear factor $\widehat{\F}_0$. The relative gains vary across different machine learning models. For instance, under the simpler models — Lasso, Ridge, and RF — the neural network-based factor $\widehat{\F}_{\mathrm{fnn}}$ outperforms all other types, underscoring the strength of neural networks in capturing complex semantic information from text. 
Note that the FNN factor actually uses some information about the response variable (supervised), while all other methods do not (unsupervised). This explains why FNN outperforms and suggests improvements of other methods using supervised feature augmentation through, for example, sure independence screening of \cite{fan2008sure}.
This advantage diminishes under more sophisticated models like GBT and FNN, where other factor types, nevertheless, achieve superior performance. This observation suggests that while some of the latent information captured by $\widehat{\F}_{\mathrm{fnn}}$ may be more or less redundant with what is learned by complex models, other augmented factors encode additional predictive structure not readily extracted by the base learners. Notably, across all five models, the Gaussian kernel factor $\widehat{\F}_{\mathrm{rbf}}$ consistently ranks among the top performers, highlighting the effectiveness of extracting nonlinear latent representations through kernel transformations for this problem.

The proposed augmentation method proves useful not only in regression settings but also in classification tasks, such as sentiment estimation. In the following two subsections, we further examine the stock market by defining sentiment scores, which are then used to conduct event studies and portfolio analyses. These evaluations serve as supplementary assessments of the estimation performance of the proposed feature augmentation methods. The goal of these experiments is to evaluate the effectiveness of the augmentation strategies when applied to news text data in the context of financial investment. To this end, we compare the performance of estimations based on factors derived from various transformations, following the financial market analysis framework established in \citet{fan2021much}.


\subsection{Event Study}
\begin{figure*}[!htb]
	\begin{center}
		\captionsetup{width=\linewidth}
		\subfigure{\includegraphics[width=0.48\textwidth]{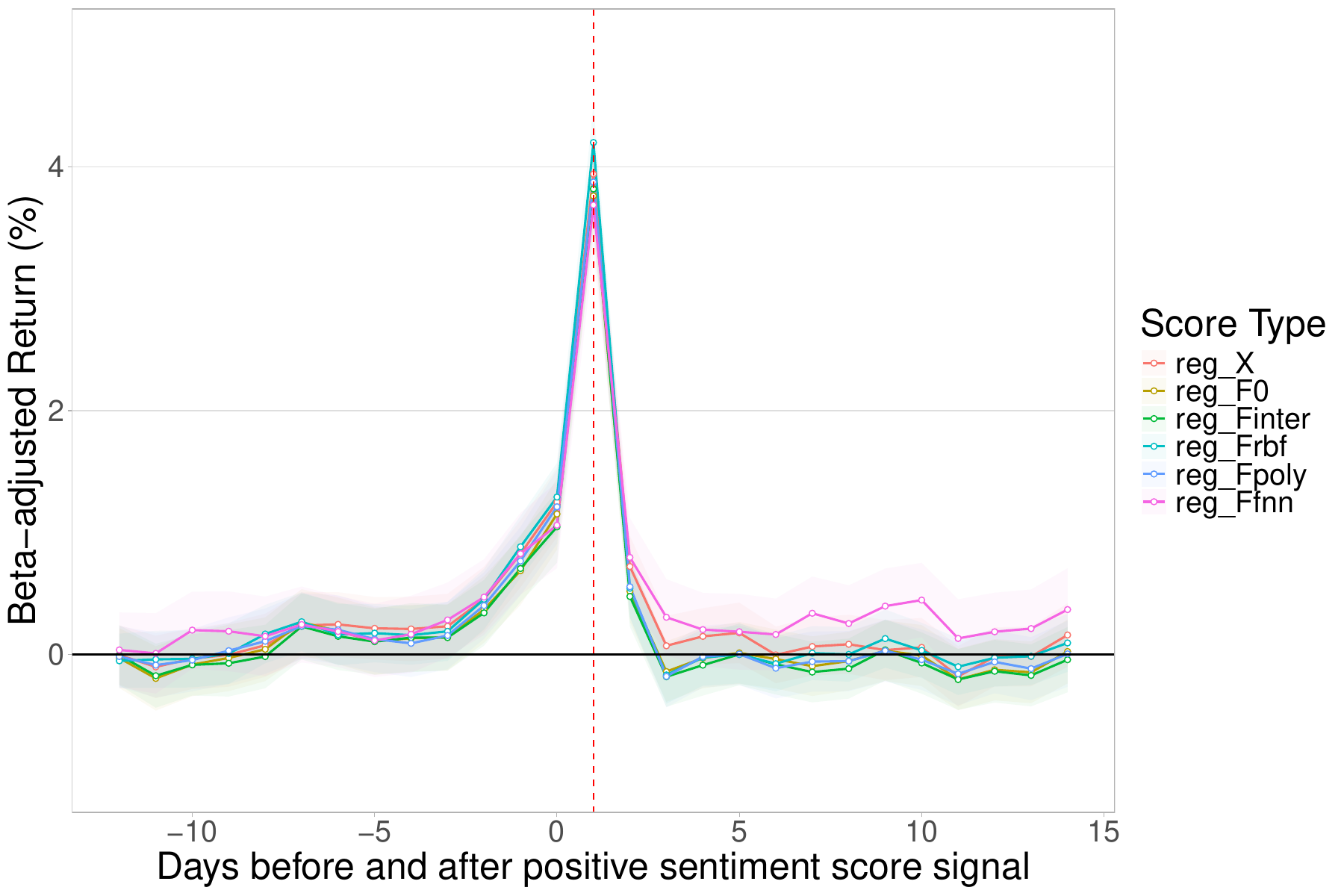}}
		\hspace{0.3cm}
		\subfigure{\includegraphics[width=0.48\textwidth]{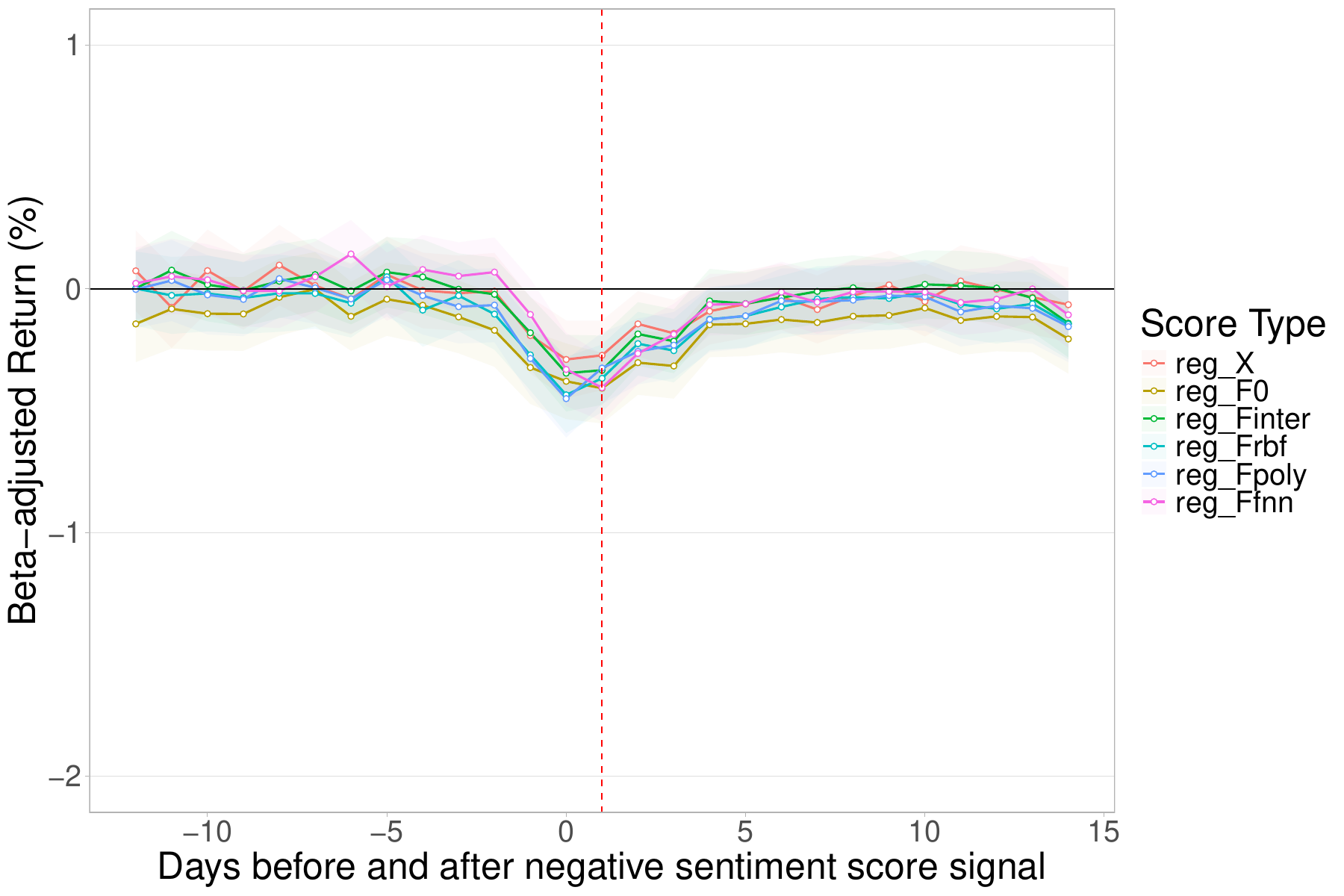}}

		\vspace{0.3cm}  

		\subfigure{\includegraphics[width=0.48\textwidth]{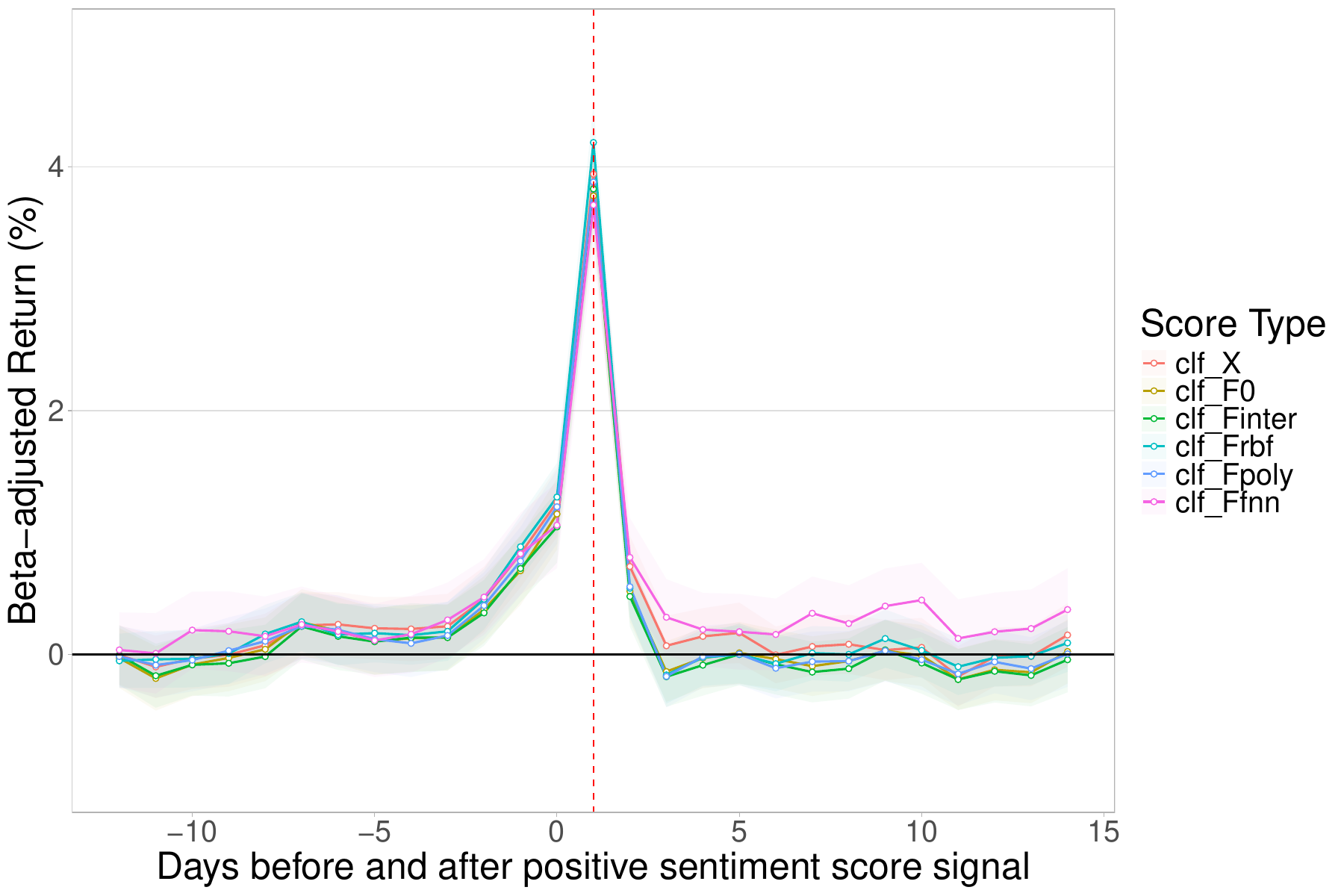}}
		\hspace{0.3cm}
		\subfigure{\includegraphics[width=0.48\textwidth]{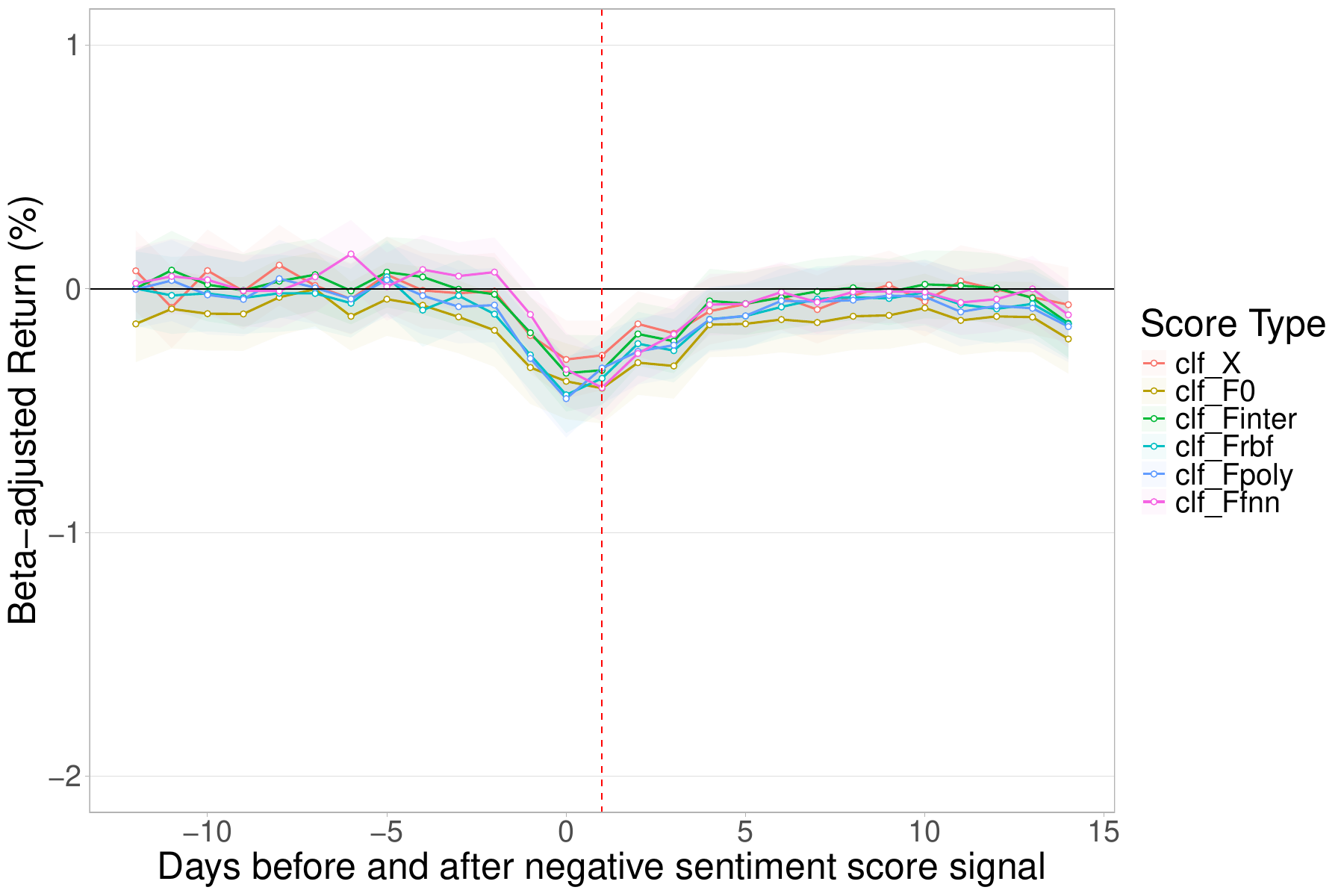}}

		\setlength{\belowcaptionskip}{-20pt}
		\setlength{\abovecaptionskip}{0pt}
		\caption{Event Study on Beta-Adjusted Returns. The $x$-axis represents the number of days $p$ before and after the signal (news publication), and the $y$-axis shows the estimated $\beta_{i,p}$. We set day 1 as the day of the event occurring. The curves depict the average response across all stocks, with shaded areas indicating $95\%$ confidence intervals. The two figures on the top show the results for the regression estimators, and the two at the bottom are for the binary classification scores. 
        The red, brown, green, blue, cyan, and pink curves are results estimated using $\X$, $(\widehat\F_0, \widehat\U)$, $(\widehat\F_{\mathrm{fnn}}, \widehat\F_0, \widetilde{\U})$, $(\widehat\F_{\mathrm{inter}}, \widehat\F_0, \widetilde{\U})$, $(\widehat\F_{\mathrm{rbf}}, \widehat\F_0, \widetilde{\U})$, and $(\widehat\F_{\mathrm{poly}}, \widehat\F_0, \widetilde{\U})$, respectively.
}
		\label{Fig_eventstudy}
	\end{center}
\end{figure*}
\vspace{0.1in}

Recall that our dependent variable is the beta-adjusted return of a stock on the day of the associated news publication. The sign of the return can serve as a proxy for the sentiment conveyed by the news regarding the corresponding stock. Based on these sentiment scores, we conduct an event study to assess how individual stocks exhibit return responses over time in reaction to the identified sentiment weeks before and after the event. To illustrate the flexibility of the proposed feature augmentation framework, we present two approaches for generating sentiment scores, both leveraging the augmented features.

First, as described in the previous subsection, we have estimated continuous-valued returns using regressions based on different feature sets. To ensure consistency across rolling windows, we re-center the predicted values by subtracting the mean of the training data (which is very close to zero) and adding 0.5, so that the output remains interpretable as a centered score. Due to space constraints, we report results only for the score estimations obtained using Random Forest.
Alternatively, since the event study focuses on the direction of stock returns, we can adopt a classification-based approach that simplifies the estimation task and enhances robustness. Specifically, we dichotomize the target variable and, as an example, train a three-layer feedforward neural network (FNN) with two hidden layers of widths 16 and 4, each followed by a dropout layer with a rate of 0.2. The model outputs the estimated probability of a positive return, which we interpret as a sentiment score. This approach is less sensitive to noise and outliers and aligns naturally with the binary nature of sentiment.
For both methods, we compare sentiment scores derived from six different feature sets: (i) the original feature matrix $\X$; (ii) the linear factor and its associated idiosyncratic component $(\widehat\F_0, \widehat\U)$; and (iii–vi) four sets of nonlinear factors, each combined with the linear factor and decorrelated idiosyncratic component: $(\widehat\F_{\mathrm{inter}}, \widehat\F_0, \widetilde{\U})$, $(\widehat\F_{\mathrm{rbf}}, \widehat\F_0, \widetilde{\U})$, $(\widehat\F_{\mathrm{fnn}}, \widehat\F_0, \widetilde{\U})$, and $(\widehat\F_{\mathrm{poly}}, \widehat\F_0, \widetilde{\U})$. 


Based on these sentiment scores, we conduct an event study to examine whether individual stocks exhibit significant return responses over time in reaction to the identified sentiment.
Each news publication is regarded as an “event”, with the sign of the sentiment score serving as a proxy for the sentiment of the news—positive scores indicate favorable news, while negative scores suggest adverse sentiment. The magnitude of the scores indicates the extent of positive or negative. We examine how individual stock returns respond to such sentiment signals. To mitigate the influence of noise, we only focus on the extreme events with the top 5\% of positive or negative scores, as determined by the absolute magnitude of their fitted returns.

For positive news, we define an event for stock $i$ on day $t$ if there are news articles about stock $i$  published between the market close (3:00 pm) of day $(t{-}1)$ and the market close of day $t$, and its associated score falls within the top 5\% quantile of all positive scores. If there is more than one news article about stock $i$ on day $t$, we take the average of the scores.
We then estimate the following event-study regression:
\begin{equation}\label{equ:event}
\mathrm{Return}_{it} = \sum_{p = -13}^{14} \beta_p \mathrm{Day}_{ip} + \delta_i + \mu_t + \varepsilon_{it},
\end{equation}
where $\mathrm{Return}_{it}$ denotes the realized beta-adjusted return of stock $i$ on day $t$; $\mathrm{Day}_{ip}$ is an indicator of $p$ days after the event; and $\delta_i$ and $\mu_t$ are stock and day fixed effects, respectively, to account for unobserved heterogeneity. An analogous regression is conducted for negative news events.

Figure~\ref{Fig_eventstudy} presents the results of the event-study regressions for positive and negative news events during the period 2015–2019. The top two panels correspond to sentiment scores obtained via Random Forest regression, while the bottom two are based on scores generated through binary classification using an FNN. The findings are consistent across both methods.
Scores derived from feature augmentation yield sharper event curves compared to the baseline, where scores are computed directly from the original feature matrix $\X$ without augmentation. This suggests that the augmented features provide a more accurate estimation of sentiment, particularly for extreme positive and negative events. For the top 5\% of positive sentiment cases, beta-adjusted returns—as captured by the estimated coefficients $\beta_p$—begin to rise approximately seven days before the event, accelerate two days prior, and peak on the day of the news release.
One possible cause is that, due to possible new leakage or anticipation, investors start to buy stocks before the positive news.
In contrast, negative sentiment events trigger return responses that are more concentrated around the event day, with generally lower magnitudes. This asymmetry suggests that negative news has a more limited impact on stock prices than positive news in the Chinese market. This is a consequence of short-sale restrictions, which limit investors' ability to profit from negative information. 
Even if investors got negative news in the Chinese market, they can not easily make profits due to the short constraints.  Shareholders gradually hear the news and add selling pressure a few days after the event.
This behavior contrasts with that observed in more liberalized markets such as the United States.



\subsection{Portfolio Performance}
We further construct stock portfolios based on the scores from 2015 to 2019 and evaluate the effectiveness of the proposed feature augmentation methods. 
We compare the performance of different augmentation strategies in a practical investment setting and adopt a long-short portfolio strategy: among news published each day, buy the top 50 stocks with the highest sentiment scores and sell the bottom 50 with the lowest scores. If fewer than 50 stocks have scores larger (or smaller) than 0.5 on a given day, the unallocated capital is held as cash and earns no interest. We consider the value-weighted (VW) strategies, where portfolio weights are proportional to the stocks’ total market capitalization on the preceding trading day.
Additionally, we account for the daily transaction costs in the Chinese equity market. Transaction costs are applied each day when the portfolio changes. These include stamp duty, transfer fees, and trading commissions, following typical practices in the Chinese stock market. We assume a total transaction cost of 13 basis points for each round-trip trade (buy and sell). Portfolio returns are adjusted for turnover to reflect trading costs accurately. 

\begin{figure*}[!htb]
	\begin{center}
		\captionsetup{width=\linewidth}

		\vspace{0.3cm}  

		\subfigure{\includegraphics[width=0.48\textwidth]{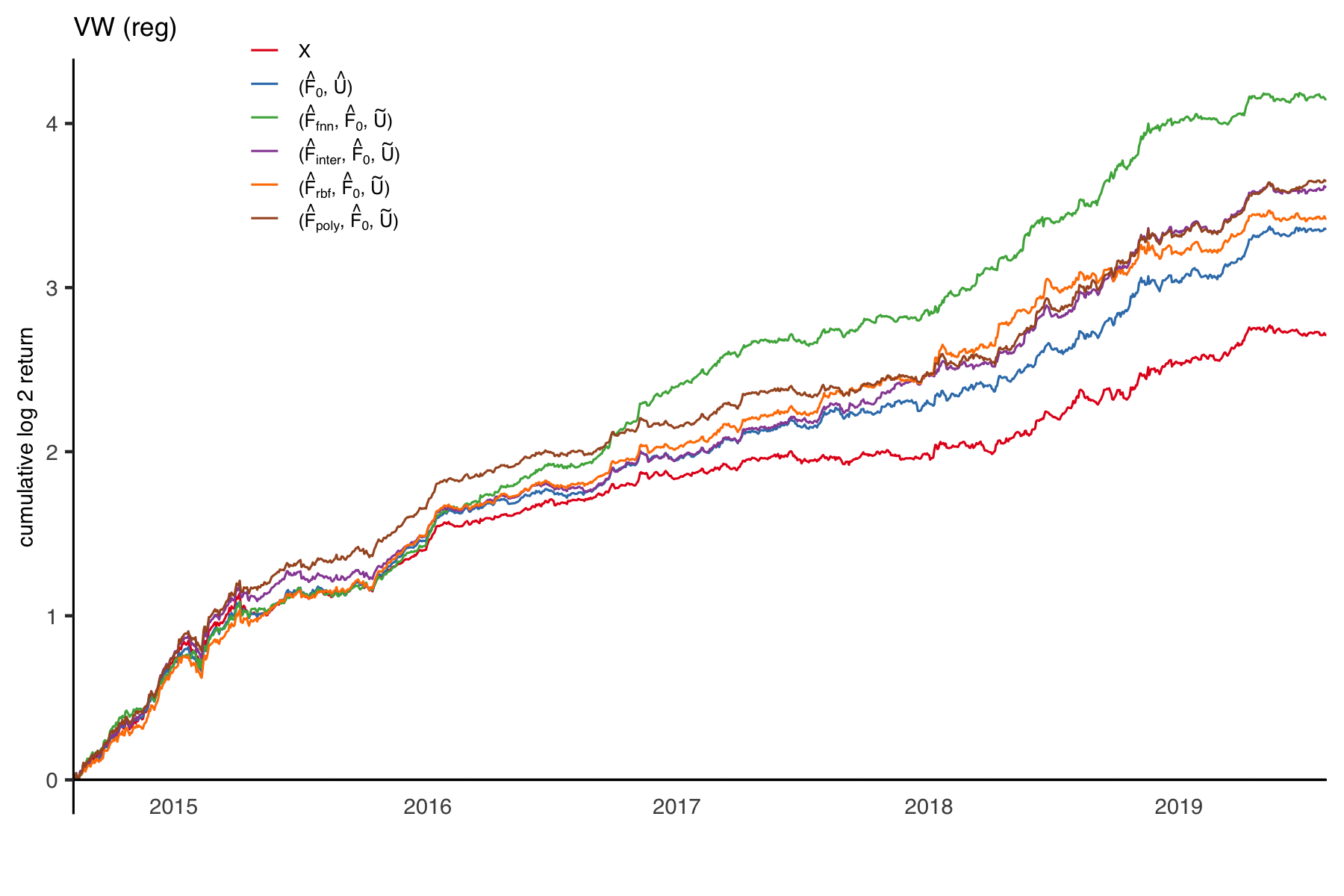}}
		\hspace{0.3cm}
		\subfigure{\includegraphics[width=0.48\textwidth]{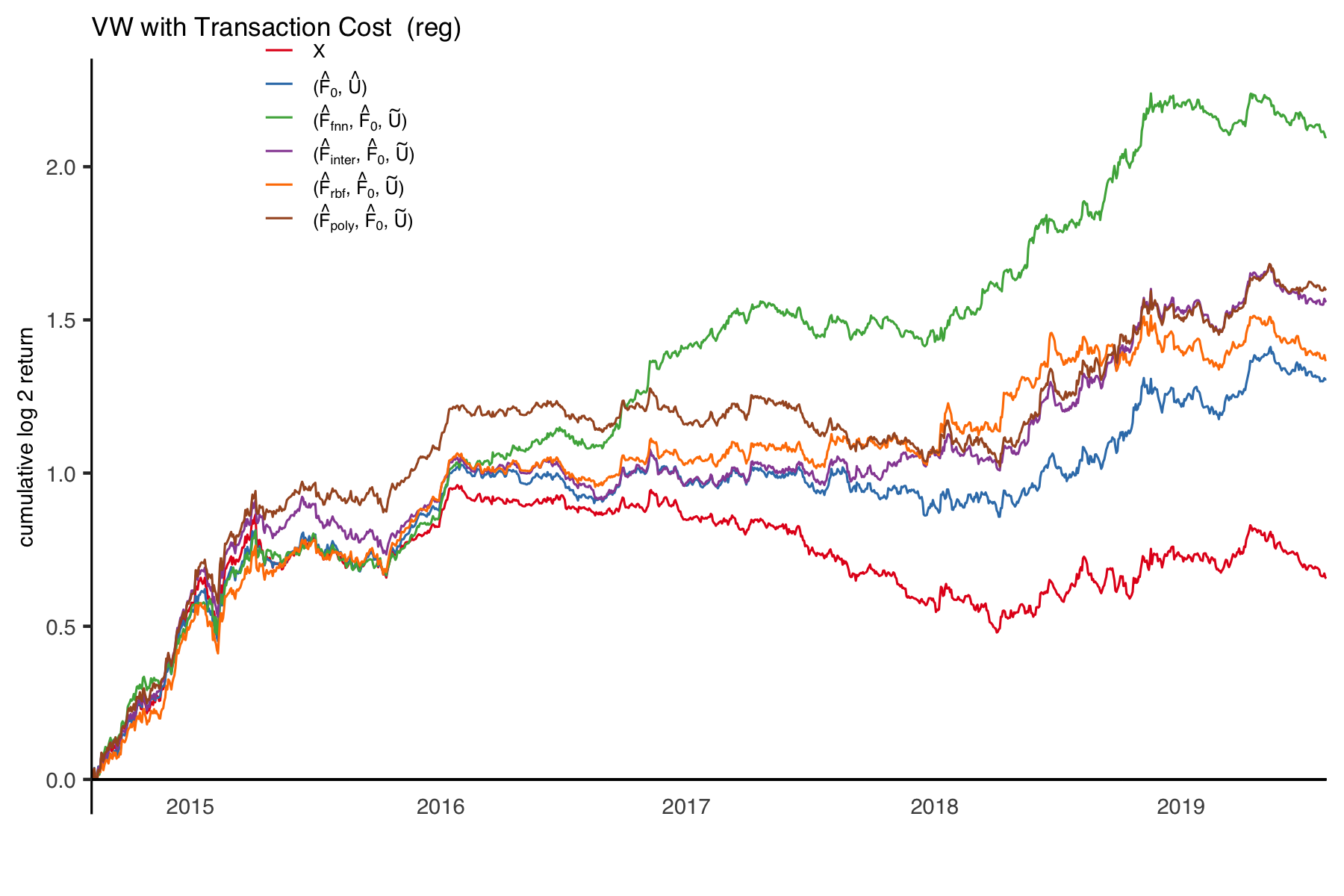}}




		\subfigure{\includegraphics[width=0.48\textwidth]{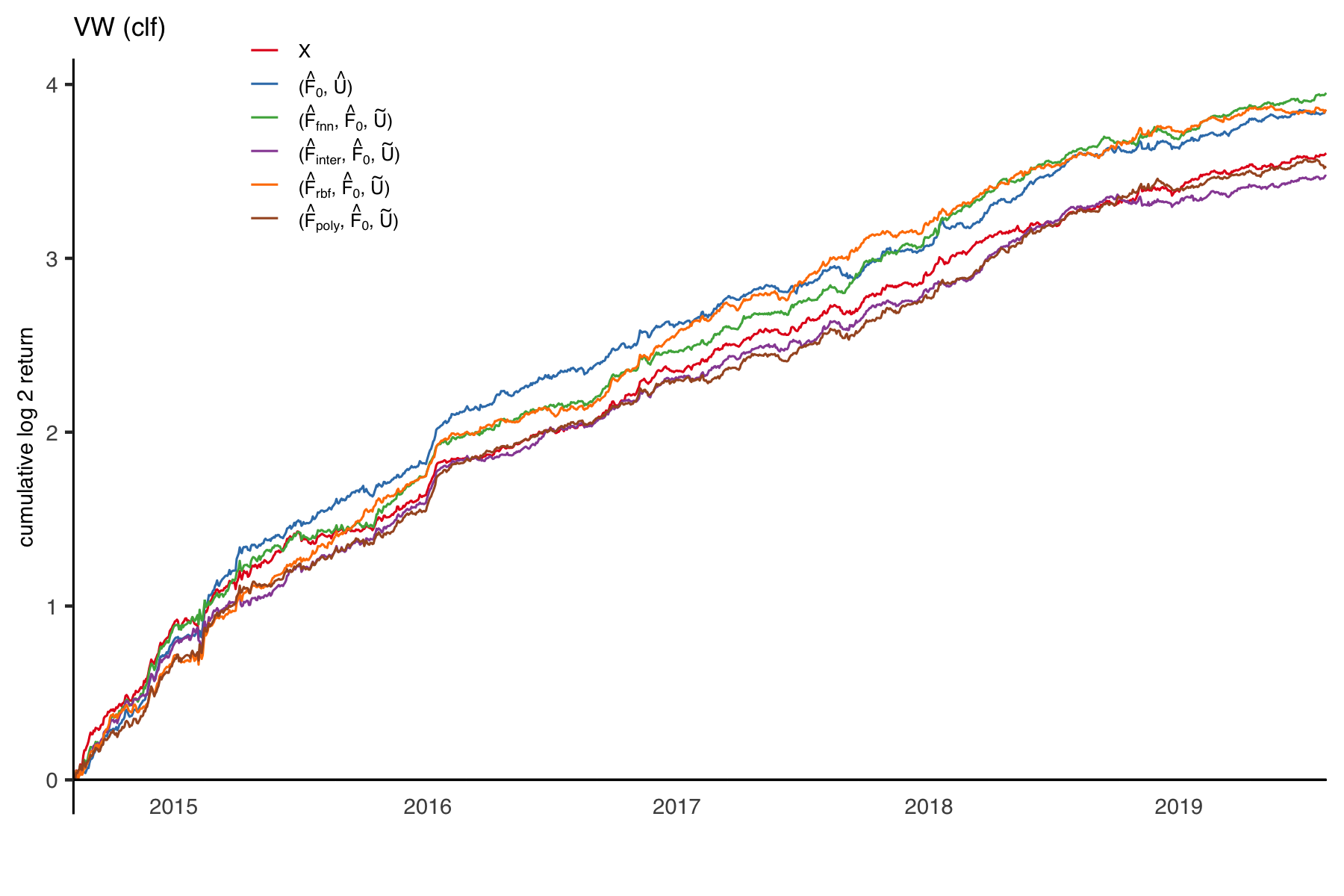}}
		\hspace{0.3cm}
		\subfigure{\includegraphics[width=0.48\textwidth]{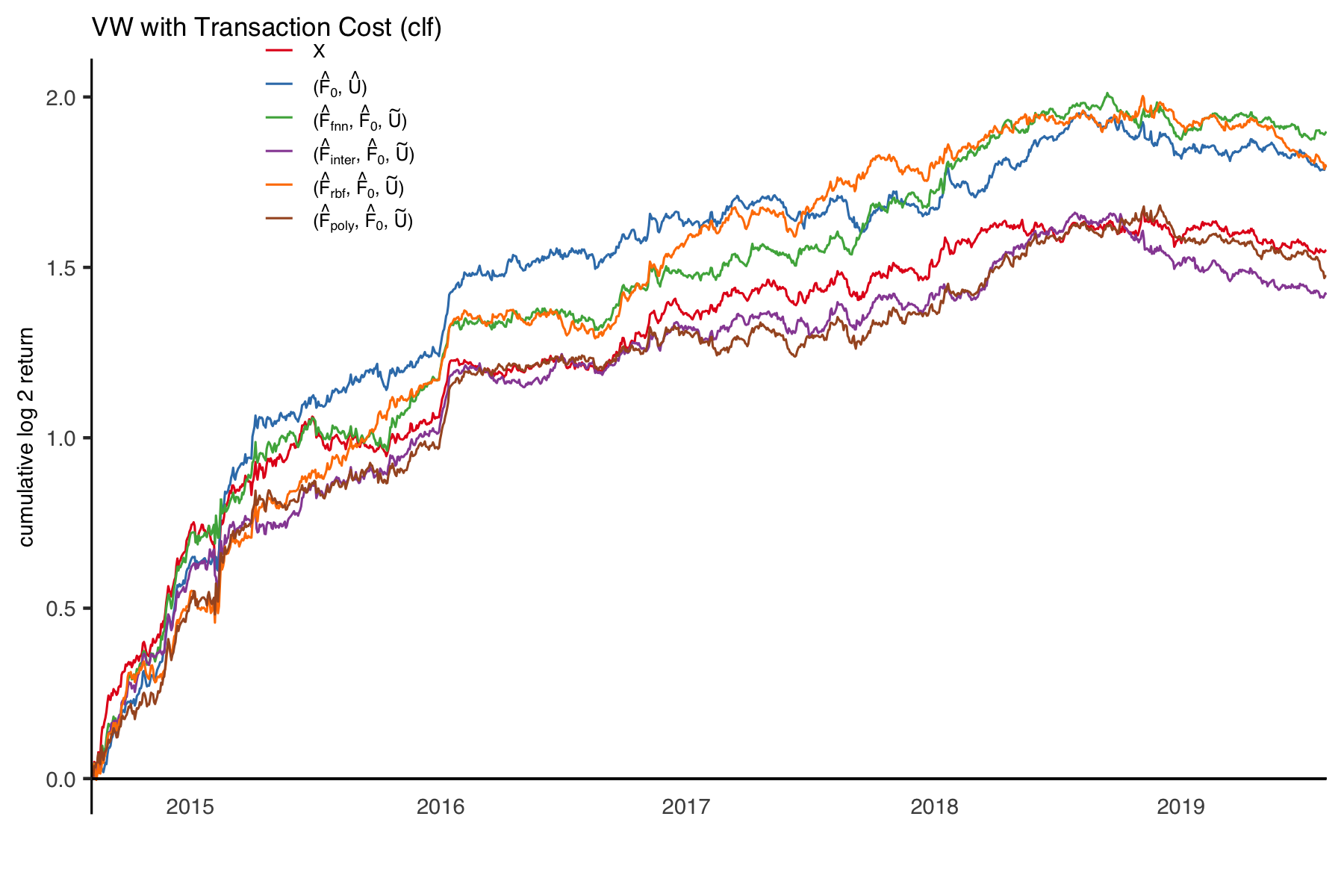}}

		\setlength{\belowcaptionskip}{-20pt}
		\setlength{\abovecaptionskip}{0pt}
		\caption{Cumulative Log$_2$ Returns of Value-Weighted (VW) Portfolios. The top two panels display results based on regression-based score estimators, while the bottom two panels correspond to scores derived from binary classification.
        The red, blud, green, purple, orange, and brown lines corresponds to the investment return based on scores estimated from $\X$, $(\widehat\F_0, \widehat\U)$, $(\widehat\F_{\mathrm{fnn}}, \widehat\F_0, \widetilde{\U})$, $(\widehat\F_{\mathrm{inter}}, \widehat\F_0, \widetilde{\U})$, $(\widehat\F_{\mathrm{rbf}}, \widehat\F_0, \widetilde{\U})$, and $(\widehat\F_{\mathrm{poly}}, \widehat\F_0, \widetilde{\U})$, respectively.}
		\label{Fig_portfolio_EW}
	\end{center}
\end{figure*}

Figure~\ref{Fig_portfolio_EW} presents the results of the portfolio analysis based on score estimations obtained using binary classification with FNN. The top two panels show results based on regression-based score estimators, while the bottom two correspond to scores derived from binary classification. The panels on the left display portfolio performance without trading costs, whereas the right panels incorporate transaction costs.
Across all settings, portfolios constructed using scores derived from feature augmentation consistently outperform the baseline model that uses only the original feature matrix. In particular, the factors extracted via the FNN-based transformation, $(\widehat\F_{\mathrm{fnn}}, \widehat\F_0, \widetilde{\U})$, yield the best performance, in line with the findings reported in Section~\ref{sec:chinese_result}.

\begin{table}[!htb]
\centering
\caption{Portfolio performances from 2015 to 2019 with transaction fees using VW strategies}
\label{tab:portfolio}
\resizebox{\textwidth}{!}{%
\begin{tabular}{@{}lrrrrrrrrrrrr@{}}
\toprule
          & \multicolumn{2}{c}{$(\X)$}                          & \multicolumn{2}{c}{$(\widehat\F_0, \widehat\U)$}    & \multicolumn{2}{c}{$(\widehat\F_{\mathrm{inter}}, \widehat\F_0, \widetilde{\U})$} & \multicolumn{2}{c}{$(\widehat\F_{\mathrm{rbf}}, \widehat\F_0, \widetilde{\U})$} & \multicolumn{2}{c}{$(\widehat\F_{\mathrm{poly}}, \widehat\F_0, \widetilde{\U})$} & \multicolumn{2}{c}{$(\widehat\F_{\mathrm{fnn}}, \widehat\F_0, \widetilde{\U})$} \\ \cmidrule(l){2-13} 
Portfolio & \multicolumn{1}{c}{APR \%} & \multicolumn{1}{c}{SR} & \multicolumn{1}{c}{APR \%} & \multicolumn{1}{c}{SR} & \multicolumn{1}{c}{APR \%} & \multicolumn{1}{c}{SR} & \multicolumn{1}{c}{APR \%} & \multicolumn{1}{c}{SR} & \multicolumn{1}{c}{APR \%} & \multicolumn{1}{c}{SR} & \multicolumn{1}{c}{APR \%} & \multicolumn{1}{c}{SR} \\ \midrule
Classification        &                         &                      &                         &                      &                         &                      &                         &                      &                         &                      &                         &                      \\
\quad L+S & 21.4                    & 1.89                 & 25,4                    & 2.14                 & 19.3                    & 1.71                 & 25.9                    & 2.12                 & 20.2                    & 1.75                 & 27.3                    & 2.23                 \\
\quad L   & 35.1                    & 3.39                & 36.2                    & 3.50                & 33.5                    & 3.26                & 37.8                    & 3.56                & 31.6                   & 3.16                 & 36.0                     & 3.48                 \\
\quad S   & -10.2                    & -1.80                 & -8.0                    & -1.36                 & -10.8                    & -1.83                 & -8.72                    & -1.42                & -8.74                    & -1.46                 & -6.49                    & -1.08                 \\
Regression        &                         &                      &                         &                      &                         &                      &                         &                      &                         &                      &                         &                      \\
\quad L+S & 7.28                    & 0.626                 & 17.5                    & 1.34                 & 21.7                    & 1.61                & 18.8                    & 1.37                 & 22.6                    & 1.61                 & 31.4                    & 2.14                 \\
\quad L   & 28.2                    & 2.47                & 35.6                    & 2.93                 & 39.9                     & 3.19                 & 38.6                     & 3.03                 & 40.6                     & 3.14                 & 45.4                     & 3.38                 \\
\quad S   & -16.3                    & -2.61                 & -13.3                    & -2.23                 & -13.0                    & -2.20                 & -14.2                    & -2.33                 & -12.8                    & -2.21                 & -9.65                    & -1.75                 \\ \bottomrule
\end{tabular}%
}
\vspace{1mm}
\begin{minipage}{\textwidth}
\footnotesize\textit{Notes:} This table reports the annualized percentage returns (APR) and Sharpe ratios (SR) of value-weighted portfolio strategies, accounting for transaction costs. “Classification” refers to portfolios constructed using binary classification-based scores, while “Regression” refers to those based on regression-estimated scores. “L+S” denotes the long-short strategy, “L” denotes the long-only leg, and “S” denotes the short-only leg. 
\end{minipage}
\end{table}

Table~\ref{tab:portfolio} reports the detailed performance of the combined long-short portfolio, as well as its long and short components. We focus on the setting that incorporates the transaction costs. It shows that scores estimated from augmented features yield higher annualized percentage returns (APR) and Sharpe ratios compared to the baseline. Furthermore, factors extracted from nonlinear transformations of the original design matrix enhance the informativeness of the scores and further improve portfolio performance. Consistent with the results in Figure~\ref{Fig_eventstudy}, the majority of portfolio gains are realized from the long leg rather than the short leg. Actually, the returns gained from the short legs are negative, and this is because of the way we construct the portfolio and that we consider the transaction cost.

While we consider two score estimation methods—Random Forest regression and FNN-based binary classification—the classification-based scores generally lead to superior portfolio performance. 
That is likely due to the fact that the regression-based estimates tend to be less robust and more sensitive to noise, and thus tend to underperform compared to classification-based sentiment estimates. However, we observe in Table~\ref{tab:portfolio} that regression-based scores using Random Forest with feature sets such as $(\widehat\F_{\mathrm{inter}}, \widehat\F_0, \widetilde{\U})$, $(\widehat\F_{\mathrm{poly}}, \widehat\F_0, \widetilde{\U})$, and $(\widehat\F_{\mathrm{fnn}}, \widehat\F_0, \widetilde{\U})$ actually outperform their classification-based counterparts. This highlights the value of incorporating nonlinear feature transformations, which can significantly strengthen the predictive capacity of the feature set.

\section{More Empirical Analysis on Diverse Datasets} \label{sec_empirical}
We emphasize that the proposed feature augmentation framework is broadly applicable across diverse problem domains and learning algorithms. To demonstrate its general effectiveness in improving estimation performance, we complement the main study on Chinese news text data with a series of extensive experiments spanning both classification and regression tasks. These tasks are motivated by real-world applications in image recognition, biology, finance, etc. For each example and each learning method, we evaluate the performance of the proposed methods and compare them to those without feature augmentations. Besides, typical performances of various factors across datasets and methods are also analyzed. The objective of these experiments is twofold: to assess whether feature augmentation consistently enhances estimation accuracy across settings, and to gain better insights into which types of factors perform best under different data structures and learning contexts.

To accommodate the characteristics of these auxiliary datasets, we introduce several adjustments to the experimental setup. For image-based problems, we incorporate the factor $\F_{\mathrm{cnn}}$, extracted from the last hidden layer of a convolutional neural network (CNN) applied to the original feature matrix $\X$, capitalizing on CNNs' proven capabilities in visual representation learning. Additionally, we employ task-appropriate error metrics for classification and non-temporal regression settings. Since most of these datasets are substantially smaller than the Chinese news corpus, we are able to utilize finer cross-validation grids for hyperparameter tuning, which we keep fixed across all settings for clarity.  We consider all five machine learning techniques (Lasso, Ridge, RF, GBT, and FNN). Since Neural Networks are generally not easy to tune for regression problems if the input data is structured (organized in rows and columns with clearly defined, curated features) \citep{grinsztajn2022tree, shwartz2022tabular}, only the first four methods are applied to the regression datasets.

Moreover, another interesting idea considered here is to explore the potential of combining different augmentation methods to further enhance model performance. Two approaches are proposed for this purpose: aggregating different factors and combining factors with other feature augmentation techniques.
We have already illustrated and used the first approach in the previous section, and we will do the same for all the rest datasets.
The second approach is adopted mainly in binary classification problems in which log-likelihood ratios are considered as another augmentation method \citep{fan2016feature}. Let $f_{1j},\ f_{2j}$ be the densities of $j^{th}$ feature in class 1 and 2.
The log-likelihood ratio for the $j^{th}$ feature is defined as $\operatorname{log}\frac{f_{2j}(s_j)}{f_{1j}(s_j)}$, for $j=1,...,p$, and it is the best classifier if only $j^{th}$ feature is used.  As pointed out by \cite{fan2016feature}, naive Bayes uses the summation of these features without any training, and inputting them in the training algorithms usually leads to improvements.
The densities can be estimated by kernel density estimation using the training sample, and for the stability of the created features, we set an estimated marginal density to above some threshold $\epsilon$ (say $10^{-2}$) if it is less than $\epsilon$.

In the remainder of this section, we first describe the modified tuning and evaluation procedures, then detail the datasets and associated tasks, and finally present the empirical results — alongside those from the Chinese news data — to further validate the efficacy and adaptability of the proposed augmentation strategies.

\subsection{Tuning and Testing}
In the cross-validation for hyperparameters, we keep the maximum number of iterations to converge in Lasso and Ridge, the number of trees in RF, and the number of boosting rounds in GBT to be 100.
For all the problems, the tuning parameter $\gamma_1$ (resp. $\gamma_2$) for Lasso (resp. Ridge) is chosen from $20$ numbers from $10^{-3}$ to $10^3$ with even space on a log scale. 
In RF, for classification problems, the maximum depth of trees is chosen from $(5,10,15,20,25,30)$, and the minimum sample number required to split an internal node is chosen from $(1,3,5,8)$; for the regression problems, the maximum depth of trees is chosen from $(3,6,9,12,15,18,21,24)$ and the minimum sample number required to split an internal node is chosen from $(1,2,4,8,16)$.
In GBT, both classification and regression have the same parameter grids. The maximum depth of trees is chosen from $(5,10,15,20,25)$, the minimum sum of instance weight needed in a child is chosen from $(1,3,5,8)$, and the learning rate is chosen from $(0.1,0.3,0.5)$.

The performance of classification is measured by the classification error in the testing set, which is the rate of wrongly classified numbers to the size of the testing set.
Let $\x_1^{0},\ldots,\x_{n_{\mathrm{new}}}^{0}$ be the samples in the testing sets with the corresponding true labels $y_1^{0},\ldots,y_{n_{\mathrm{new}}}^{0}$, where $n_{\mathrm{new}}$ is the sample size of the testing set. Denote the predicted labels by $\widehat{y}_1^{0},\ldots,\widehat{y}_{n_{\mathrm{new}}}^{0}$. Let $I(\cdot)$ be the indicator function. Then, the classification error (ERR) is defined as 
\[
\mathrm{ERR} = \frac{1}{n_{\mathrm{new}}} \sum_{i=1}^{n_{\mathrm{new}}} I\big(\widehat{y}_i^{0}\neq y_i^{0}\big),
\]

The performance of regression (without rolling-window) is evaluated using the out-of-sample $R^2$, which is defined as
\[
R^2 = 1-\frac{\sum_{t=1}^{n_{\mathrm{new}}}\big(y^0_t-\widehat y^0_t\big)^2}{\sum_{t=1}^{n_{\mathrm{new}}}\big(y^0_t-\bar y_t\big)^2},
\]
where $\widehat y^0_t$ is the predicted $y^0_t$, and $\bar y_t$ is the sample mean of the response values in the training set.


\subsection{Data and pre-processing}\label{section_data}
Seven classification datasets and four groups of regression datasets are studied. To start with, we introduce the basic information of the datasets, their main features, and the preprocessing procedures.

\paragraph*{Classification datasets}
\begin{enumerate}[leftmargin=0.4cm,rightmargin=0cm]
    \item The MNIST database of handwritten digits \citep{lecun1998mnist}: The MNIST database is widely used for training various image processing systems and machine learning methods. It contains 60,000 training images and 10,000 testing images, associated with 10 labels.
    The images consist of black and white numbers from 0 to 9, each contained within a $28\times 28$ pixel bounding box.
    To speed up the process, we randomly choose $20\%$ of the training and testing sets respectively in each iteration. 
    This results in a training set of size $12000 \times 784$ and a testing set of size $2000 \times 784$.
    \item The Fashion-MNIST dataset of Zalando's article images \citep{xiao2017fashion}: The scale of Fashion-MNIST is the same as that of MNIST, consisting of 60,000 training images and 10,000 testing images. 
    Each example is a black and white image assigned to one of the ten clothing labels, including T-shirt/top, Trouser, Pullover, and so on.
    In each repetition, $20\%$ samples are randomly chosen from the training and testing sets respectively,
    which leads to a $12000\times 784$ training set and a $2000\times 784$ testing set.
    \item Kaggle Cats and Dogs dataset (DogCat): This is a binary sentiment color image classification problem, consisting of images of cats and dogs. The dataset is available through TensorFlow, with 1738 corrupted images dropped.
    The original images have different sizes and shapes, so we first standardize the dimensions to $32\times 32$ pixels bounding box.
    To speed up the experiments, all images are converted to black and white. These procedures will cost the loss of information, which is acceptable in this experiment as we are focused on the comparative performance between the models. 
    Recommendation 601 from ITU-R  is used to convert the color images to grayscale, where
    \[
    L=R\times 299/1000+G\times 587/1000+B\times 114/1000.
    \]
    After preprocessing, we get a $16283\times 1024$ training set and a $6979\times 1024$ testing set.
    \item The CIFAR-10 dataset \citep{krizhevsky2009learning}: The CIFAR-10 dataset consists of 60,000 $32\times 32$ color images in 10 classes, with 6000 images per class. The 10 labels range from animals to vehicles, including airplanes, automobiles, birds, cats, and so on.
    Similar to the DogCat dataset, all the images are converted into grayscale using Recommendation 602 from ITU-R. 
    The dataset is divided into five training batches and one test batch, with the test batch containing exactly 1000 randomly selected images from each class.
    We randomly take $20\%$ of the data in each repetition, resulting in a $10000\times 1024$ training set and a $2000\times 1024$ testing set.
    \item Reuters text categorization dataset: This is a dataset of 11,228 news articles from Reuters, labeled over 46 topics. The data is originally generated by parsing and preprocessing the classic Reuters-21578 datasets and can be accessed through TensorFlow. Each article is encoded as a list of word indexes, sorted by the words' frequency of appearance in the training set.
    The data matrix is created such that elements are set to 1 if the word corresponding to the element's index appears in the article, and 0 otherwise.
    For the purposes of this experiment, we selected the top 20-300 most frequent words to create an $8082\times 280$ training set and a $2246\times 280$ testing set. 
    \item Mice Protein Expression Dataset \citep{higuera2015self}: The dataset consists of the expression levels of 77 proteins/protein modifications that produced detectable signals in the nuclear fraction of the cortex. 
    There are in total 72 mice, with 34 of them being trisomic and the rest serving as the control group. For each protein per mouse, 15 measurements are registered, which leads to a total of  $72\times15=1080$ measurements. Each measurement can be considered as an independent sample, so the dataset is $1080\times 77$ in size. The mice are divided into eight classes based on their genotype, behavior, and treatment.
    We randomly select $80\%$ of the data as the training set, with the remaining data serving as the testing set.
    \item Cervical Cancer Behavior Risk Dataset \citep{machmud2016behavior}: The dataset contains 18 attributes regarding the risk of CA cervix behavior, with class labels of 1 and 0 indicating respondents with and without CA cervix, respectively. There are 72 samples, and $75\%$ of the data are randomly selected for training the model, with the remaining samples set as the testing set.
    Given the limited sample size, it is deemed unreliable and unnecessary to use the diversified projection method. As such, only the PCA method is used to estimate the factors.
\end{enumerate}
For the Reuters dataset and all the image classification datasets, in addition to the augmented factors,
we use $100$ screened features to reduce the computational complexity.

\paragraph*{Regression datasets} 
\begin{enumerate}[leftmargin=0.4cm,rightmargin=0cm]
    \item Prediction of the U.S. bond risk premia: The response variable is the monthly risk premia in U.S. government bonds with maturities of 2 to 5 years between January 1980 and December 2003, containing 288 data points. The $t$-year risk premium is calculated as the excess log return as the subtraction of the log holding period return of a $t$-year bond and the log yield of the 1-year discount bond \citep{cochrane2005bond}. The dataset of 2 to 5-year discount bond prices is available from the supplement for \cite{cochrane2005bond}.
    The covariates are the 127 monthly U.S. macroeconomic variables in the FRED-MD database \citep{mccracken2016fred}, which contains observations that mimic the coverage of datasets used in the literature.
    The covariates are highly correlated, as shown by \cite{fan2020factor}.
    The FRED-MD database contains {\sl NA}s due to data availability. To maintain consistency and reliability, columns containing more than $5\%$ {\sl NA}s are dropped, and the remaining 122 features are used. We apply a one-month ahead rolling window prediction with a window size of 240 months. 
    In each rolling window, we scale the data and perform CV for all algorithms.
    
    \item Prediction of taxi demand near Terminal 5, New York: The dataset is made publicly available by \cite{rodrigues2019combining}and contains 16 features, including lagged observations, weather information, and information about the presence of events. 
    The response variable is the number of individual taxi pickups that took place within a bounding box of $\pm 0.003$ geographical decimal degrees around Terminal 5.
    The samples are grouped in a daily pattern, with one year of observations (2013) being used for training, two years of data (2014-2015) for validation, and data from January 2016 to June 2016 (six months) for testing.
    \item Prediction of daily new cases of COVID-19 in the UK, the US, Singapore, and Switzerland, respectively: The COVID-19 dataset and related information are provided by Our World in Data by \cite{owidcoronavirus}. We choose 10 features, including new deaths attributed to COVID-19, number of COVID-19 patients in ICUs, government response stringency index, real-time estimate of the effective reproduction rate, total number of COVID-19 vaccination doses administered, total number of people who received at least one vaccine dose and who received all doses, daily number of people receiving their first vaccine dose, total number of COVID-19 vaccination booster doses administered, and new COVID-19 vaccination doses administered.
    The data contains observations from January 21, 2021, to July 13, 2022, consisting of 504 samples.
    We carry out a one-day-ahead rolling window prediction with a window size of 365 days. Similarly to the FRED dataset, in each rolling window, we scale the data and perform CV for all algorithms.
    \item Prediction of the house prices for Zillow: Zillow is an online real estate database company that affords a lot of data about housing prices in the US.
    The covariates consist of 17 attributes, including features of the house, for example, the number of bedrooms and bathrooms, the area of the living room, the year the house was built, the area the house is located (zip code), and so on.
    The training set consists of 15129 cases, and the testing set consists of 6484 cases. 
    We first regress the housing price on their zip codes and then regress the rest 16 features on the residual. The performance is still measured based on the estimated housing price. 
    
    \noindent
    Since the training set has a size of around $10^5\sim10^6$, it is impossible to conduct PCA on the whole dataset and to estimate the factors, especially for the two kernel factors. This is one example where diversified projection comes into play and makes factor models still applicable.
\end{enumerate}

Note: If the diversified projection is used for factor estimation, we randomly take out a tiny subset of the training set to estimate the diversified projection weight matrix and the number of factors.

\subsection{Results}\label{sec_result}
This section displays the overall results of our experiments. 
Our goal is to verify the claim that simple feature augmentations can boost the performance of various learning methods across a wide range of datasets.
Comparisons between the feature augmentation methods are also presented.
In addition, we will summarize some consistent conclusions across different methods and provide guidance on how to select factors based on the nature of the problems.



\begin{figure*}
\captionsetup{width=\linewidth}
\subfigure{\includegraphics[width=0.44\textwidth]{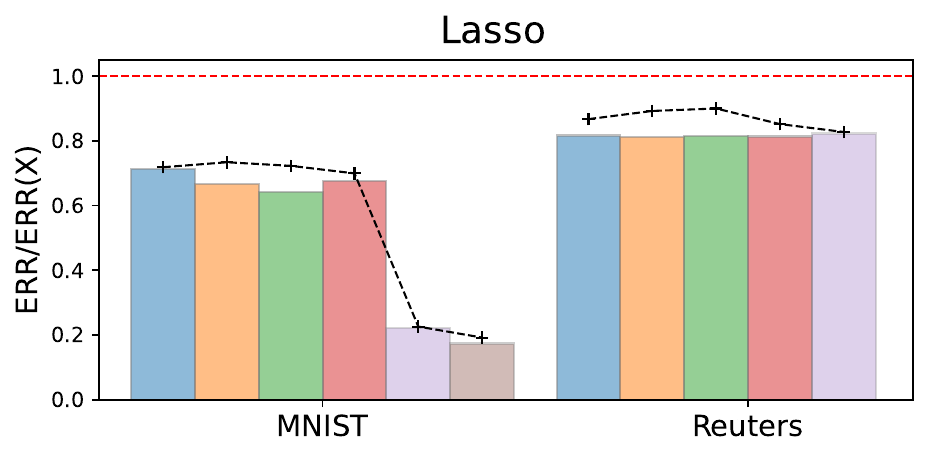}}
\subfigure{\includegraphics[width=0.44\textwidth]{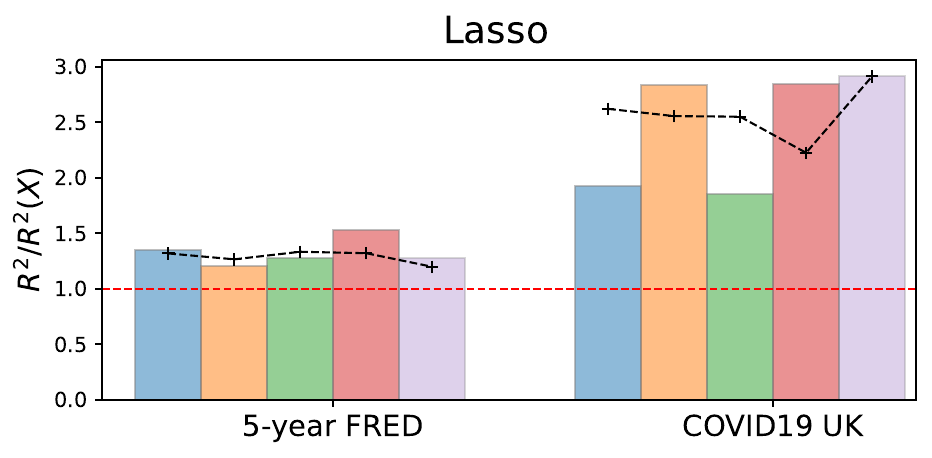}}
\subfigure{\includegraphics[width=0.44\textwidth]{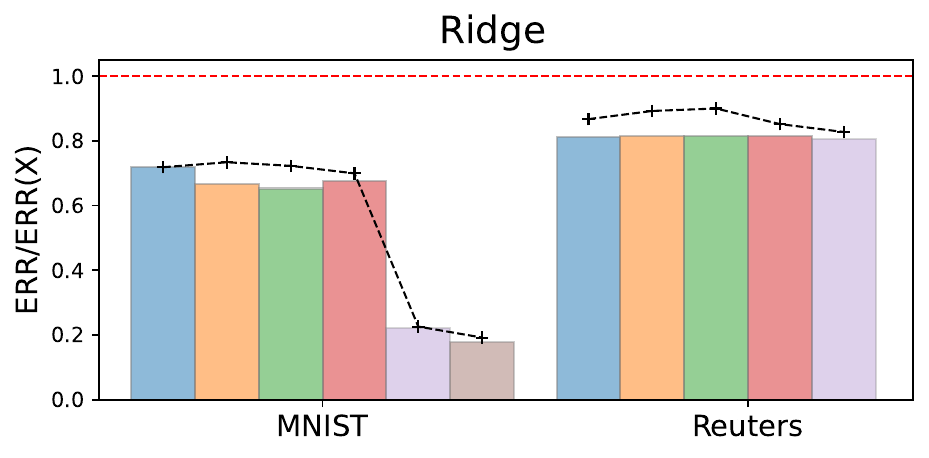}}
\subfigure{\includegraphics[width=0.44\textwidth]{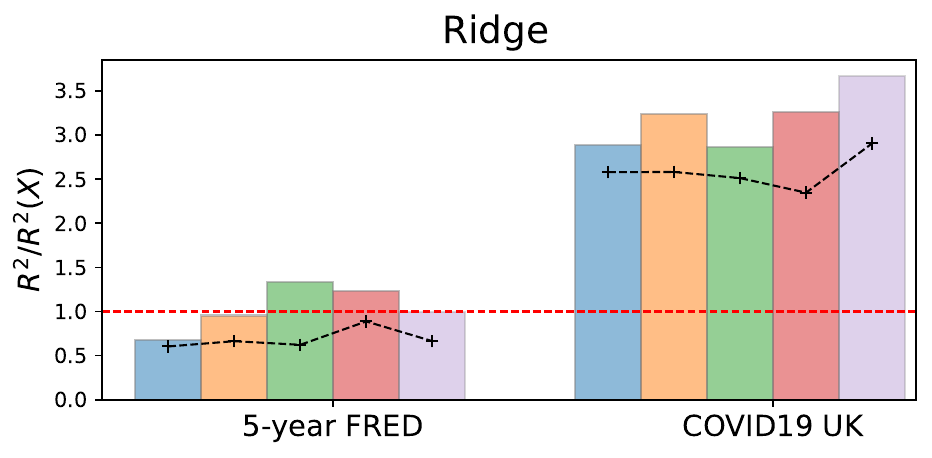}}
\subfigure{\includegraphics[width=0.44\textwidth]{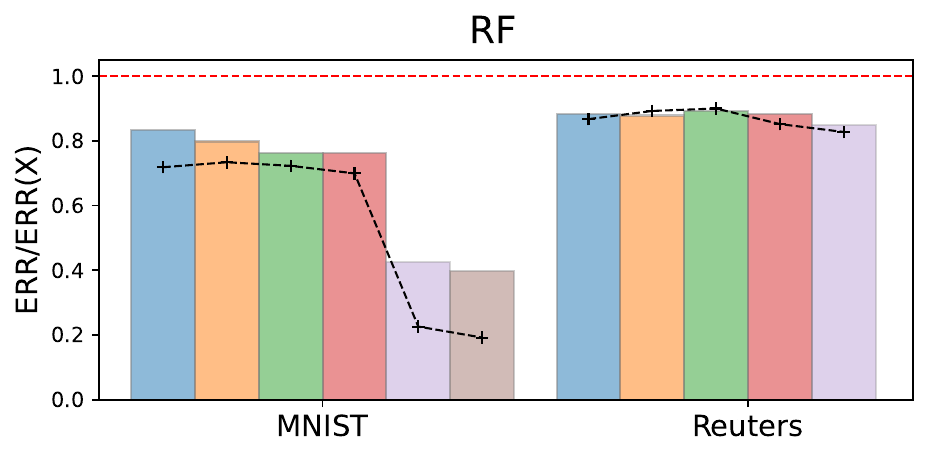}}
\subfigure{\includegraphics[width=0.44\textwidth]{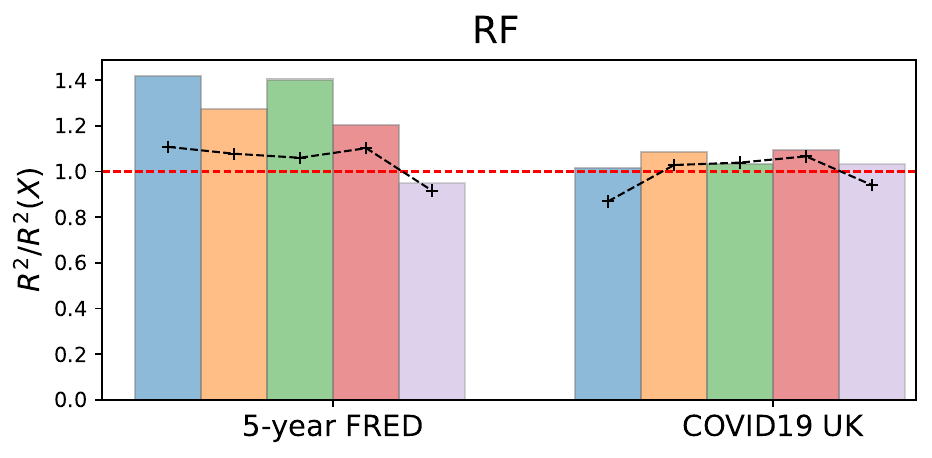}}
\subfigure{\includegraphics[width=0.44\textwidth]{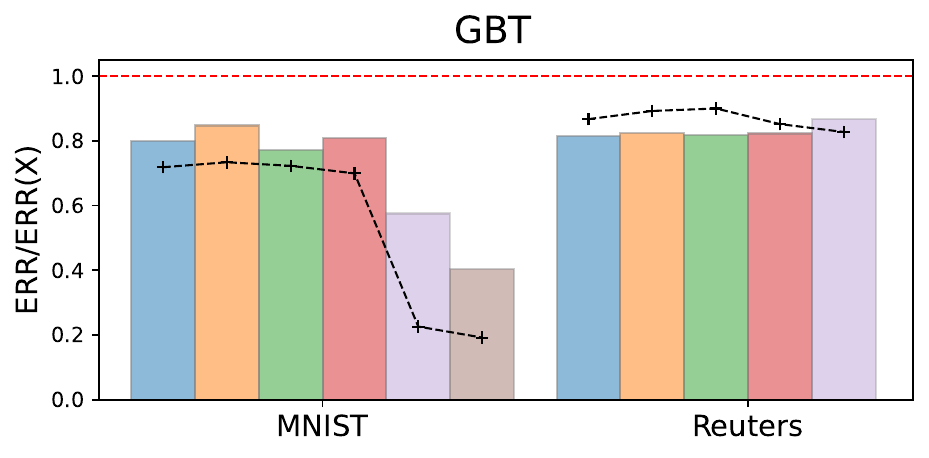}}
\subfigure{\includegraphics[width=0.44\textwidth]{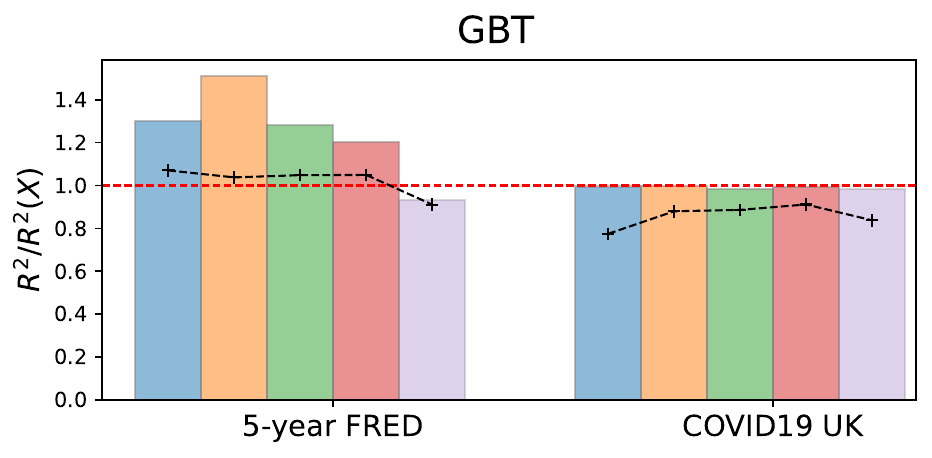}}
\subfigure{\includegraphics[width=0.44\textwidth]{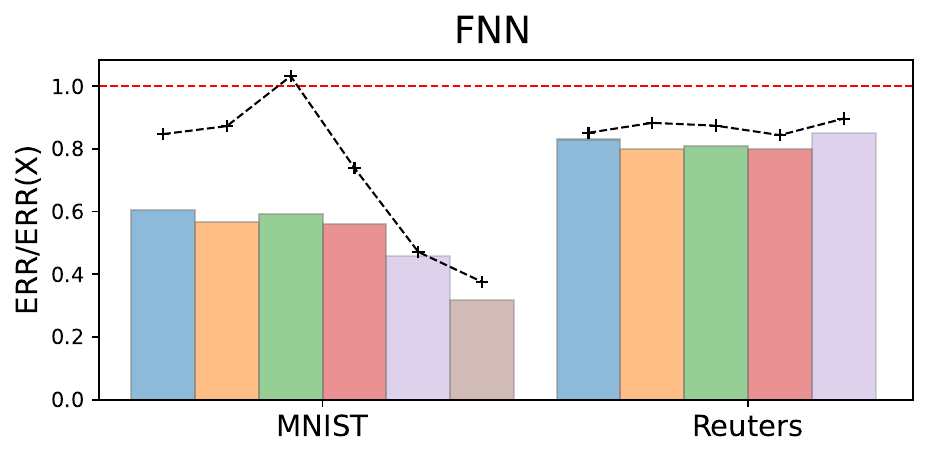}}
\hspace{4.8em}
\subfigure{\includegraphics[width=0.33\textwidth]{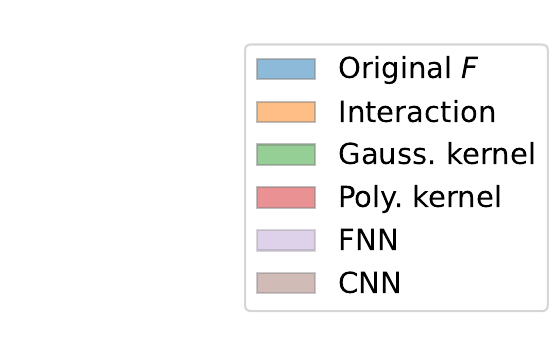}}
\caption{Left column: Ratio of the classification error (ERR) of each model to that without feature augmentation ($\mathrm{ERR}(\X)$, benchmark). The bars/points being lower than the horizontal line at 1 indicates the corresponding augmentations perform better than the benchmark.
Right column: the ratio of the out-of-sample $R^2$ of each model to that without feature augmentation ($R^2(\X)$, benchmark). The bars/points above the horizontal line at 1 indicate the corresponding augmentations perform better than the benchmark.
The histograms are results for applying PCA on the entire training set to estimate factors (cf.~Section \ref{Sec_erbd}) while the dashed lines with `+' points are the corresponding results for estimating factors by diversified projection (cf.~Section \ref{Sec_dp}). }\label{Fig_show}
\end{figure*}

\begin{figure*}[htp]
	\centering     
	\captionsetup{width=\linewidth}
	\subfigure{\includegraphics[width=0.9\textwidth]{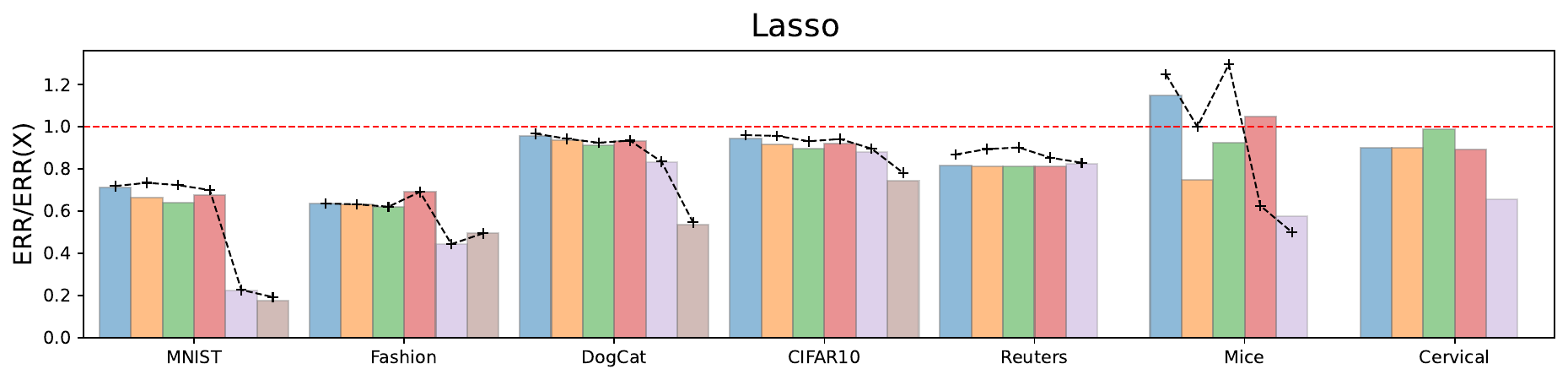}}
	\subfigure{\includegraphics[width=0.9\textwidth]{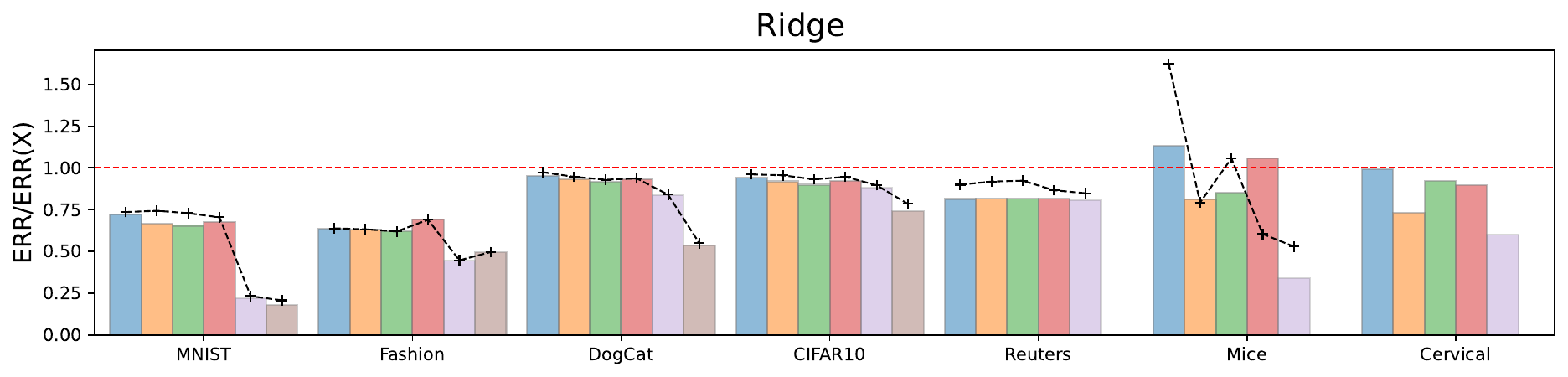}}
	\subfigure{\includegraphics[width=0.9\textwidth]{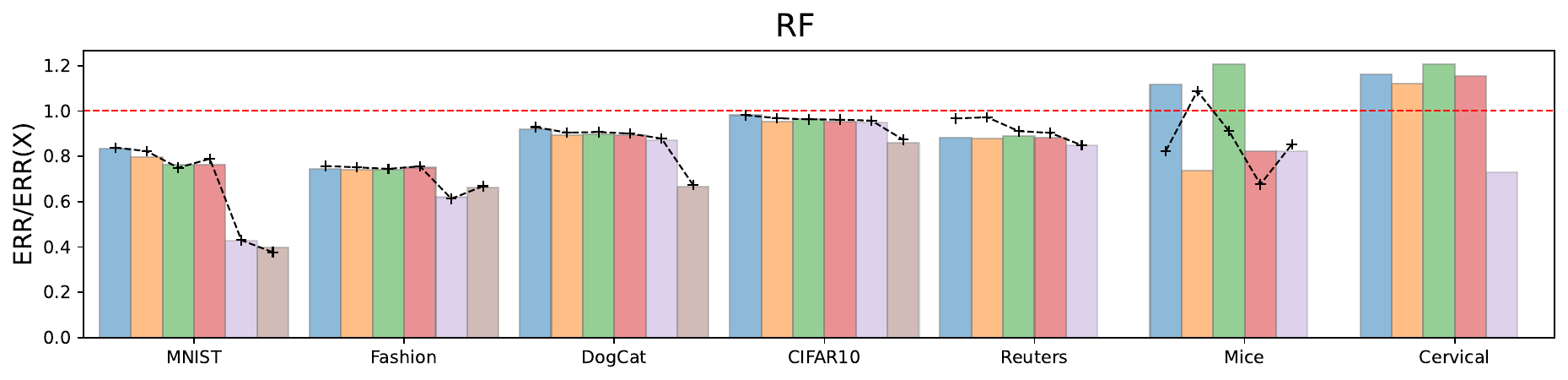}}
	\subfigure{\includegraphics[width=0.9\textwidth]{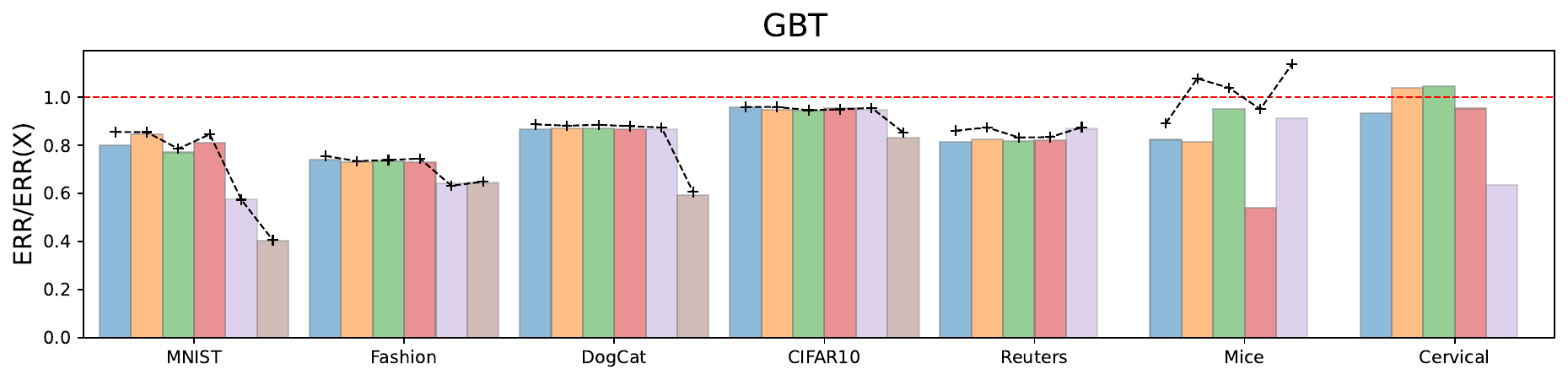}}
	\subfigure{\includegraphics[width=0.9\textwidth]{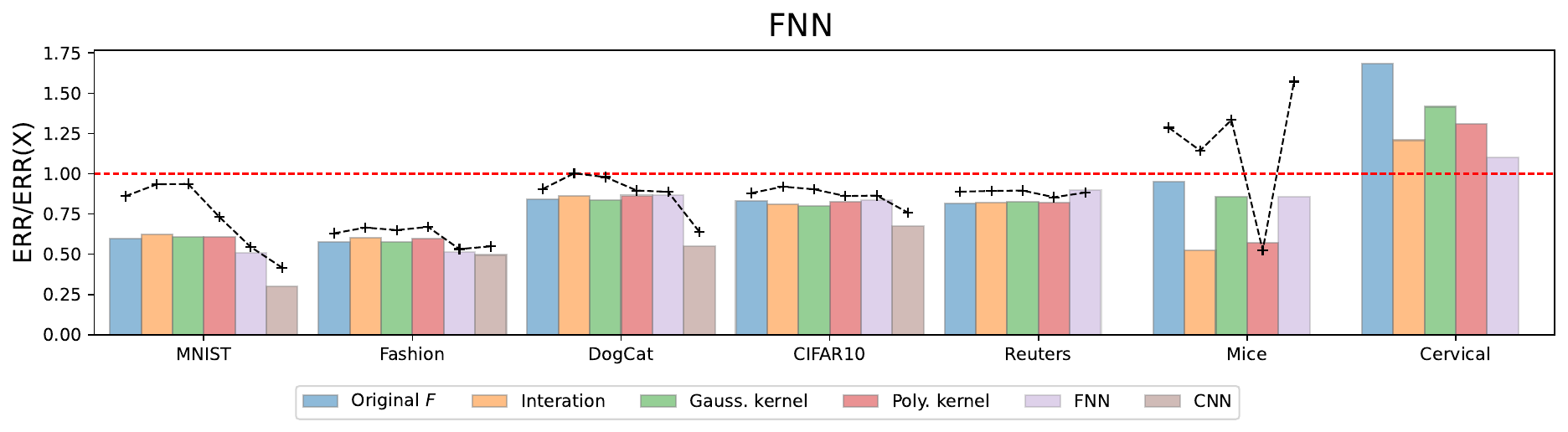}}
	\caption{Ratio of the classification error (ERR) of each model to that without feature augmentation ($\mathrm{ERR}(\X)$, benchmark). The CNN factor $\widehat{\F}_{\mathrm{cnn}}$ is only available for the images.
		The histograms are results for applying PCA on the whole data to estimate factors (cf.~Section \ref{Sec_erbd}) while the dashed lines with `+' points are the corresponding results for estimating factors by diversified projection (cf.~Section \ref{Sec_dp}). The bars/points being lower than the horizontal line at 1 indicates the corresponding factor augmentation methods perform better than the benchmark.}\label{Fig_image}
\end{figure*}
\begin{figure*}[htb]
	\captionsetup{width=\linewidth}
	\subfigure{\includegraphics[width=1\textwidth]{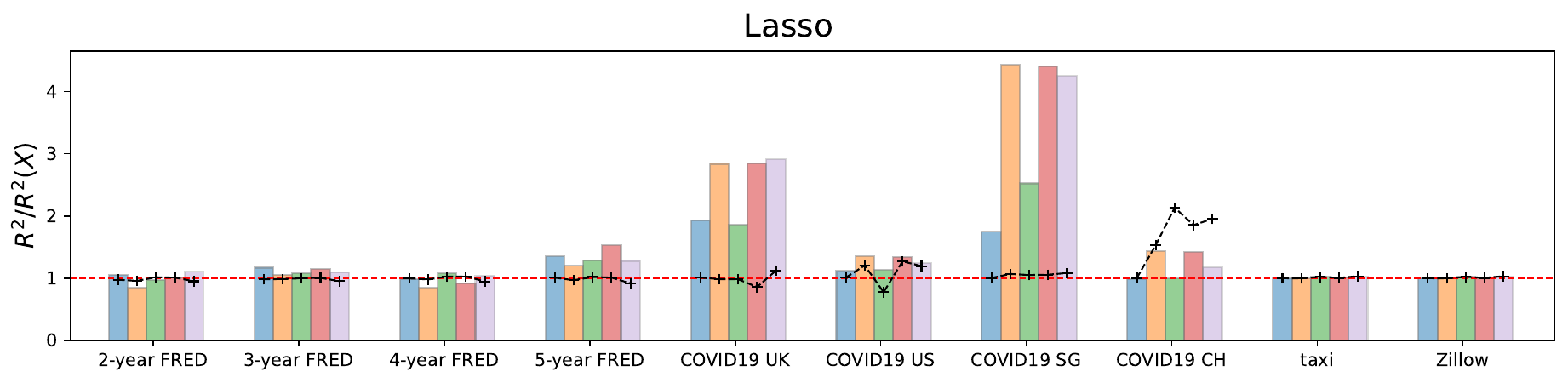}}
	\subfigure{\includegraphics[width=1\textwidth]{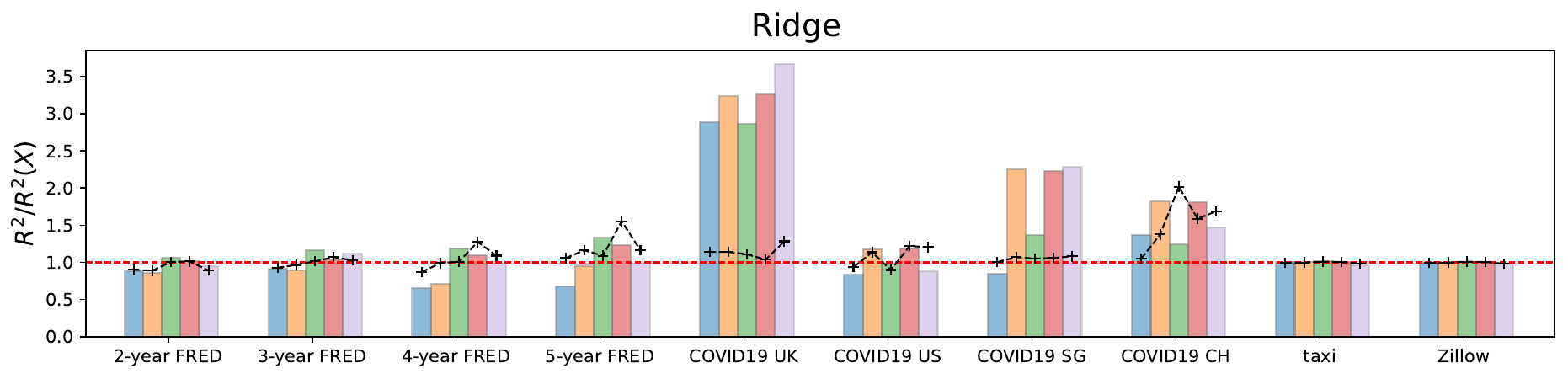}}
	\subfigure{\includegraphics[width=1\textwidth]{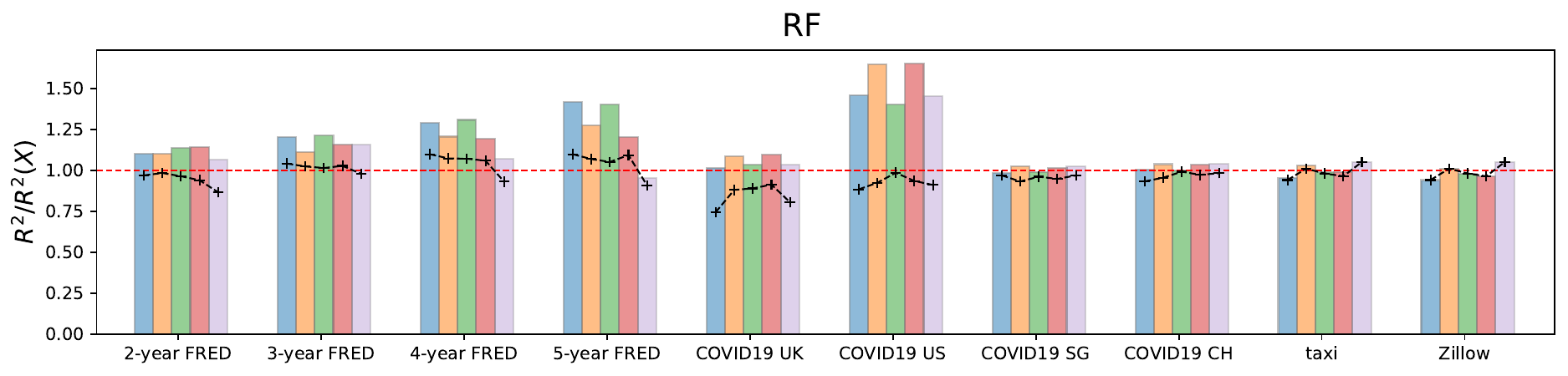}}
	\subfigure{\includegraphics[width=1\textwidth]{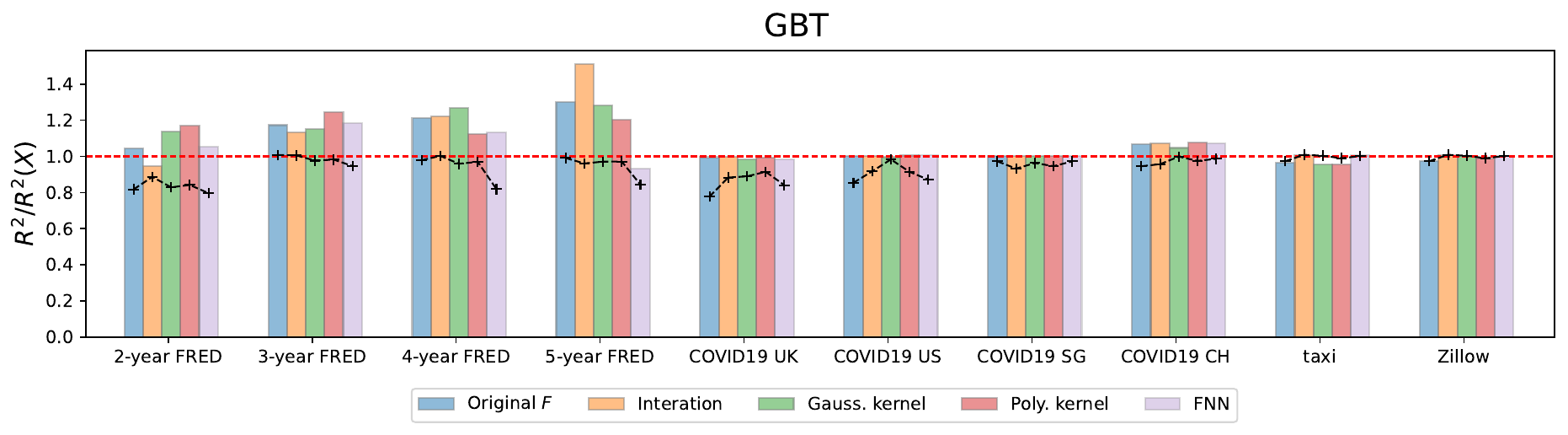}}
	\caption{Ratio of the out-of-sample $R^2$ of each model to that without feature augmentation ($R^2(\X)$, benchmark).
		The histograms are results for applying PCA on the whole data to estimate factors (cf.~Section \ref{Sec_erbd}) while the dashed lines are the corresponding results for estimating factors by diversified projection (cf.~Section \ref{Sec_dp}). The bars/points being greater than the horizontal line at 1 indicate the corresponding factor augmentation methods perform better than the benchmark.}\label{Fig_reg}
\end{figure*}

All experiments are repeated 20 times to reduce the influence of randomness, and the averages are reported.
Figure \ref{Fig_show} demonstrates the relative performance of the feature augmentation methods. Plots on the left column are for classifications while those on the right are for regressions. Only two representative results for each kind of problem are shown, and the full images can be found in Figure \ref{Fig_image} and Figure \ref{Fig_reg} at the end. All the classification errors are divided by $\mathrm{ERR}(\X)$, the classification error without feature augmentation. Smaller values indicate better performance.  
Similarly, all the out-of-sample $R^2$ are divided by $R^2(\X)$, the out-of-sample $R^2$ of the benchmark model. Greater values indicate better performance.
The histograms are results for applying PCA on the whole dataset to estimate factors (cf.~Section \ref{Sec_erbd}) while the dashed lines with `+' points are the corresponding results for estimating factors by diversified projection (cf.~Section \ref{Sec_dp}).
The blue bars are for $(\widehat{\F}_0,\widehat{\U})$, and for each of the newly proposed factors $\widehat{\F}$, we present the best outcomes among the three augmentations $(\widehat{\F},\widehat{\U})$, $(\widehat{\F}_0,\widehat{\F},\widehat{\U})$, and $(\widehat{\F}_0,\widehat{\F},\widetilde{\U})$. 

\paragraph*{Strength of Factor Models}
Factor augmentations have been shown to provide significant improvements over direct estimation using $\X$ in most cases, particularly for image classification tasks, which is evident in Figure \ref{Fig_show} and Figure \ref{Fig_image} (at the end), where almost all the bars and points are lower than $1$, indicating improved classification performance with factor augmentations. Although the results for regressions are not as consistent across algorithms and datasets as classifications, for each dataset and algorithm, there exist augmentation methods that improve the estimation. Thus, broadly speaking, factor augmentation can provide performance boosts for both types of problems. 

Besides the performance benefits over the original model without feature augmentation, we find that the latent factors obtained from transformed matrices also provide additional insights beyond $\F_0$. For almost all datasets and under all models, there always exist some transformed factors that beat the model augmented with only $\F_0$, showing the additional prediction power of extracted features from nonlinear transformations.
Below are additional comments on the performance of specific factors and datasets. 

\begin{itemize}[leftmargin=0.4cm,rightmargin=0cm]
\item As has been shown by Figure \ref{Fig_image}, for classifications, factors extracted from neural networks, $\widehat{\F}_{\mathrm{fnn}}$ and $\widehat{\F}_{\mathrm{cnn}}$, are the most powerful ones. For example, in the MNIST database using Lasso, while directly estimating with $\X$ has a high testing error of around 0.16, most factor models decrease the errors to around 0.1. Furthermore, $\widehat{\F}_{\mathrm{fnn}}$ and $\widehat{\F}_{\mathrm{cnn}}$ decrease the errors to lower than 0.04. This phenomenon can be attributed to the success of neural networks on unstructured data (e.g. raw images, audio signals, and text sequences).

\noindent In contrast, FNN factors do not always stand out prominently in regressions. Two possible reasons for this are (a) the training set is too small to train a reasonable neural network and (b) there are too many parameters to tune on neural networks. 

\item The interaction factor and kernel factors also sometimes beat $\widehat{\F}_0$ for both classification and regression problems.
Specifically, for regressions, our results first confirm the conclusion drawn in \cite{fan2020factor} regarding the bond risk premia prediction experiments that $(\widehat{\F}_0,\widehat{\U}_0)$ outperforms $\X$ under Lasso. Moreover, we find that $\widehat{\F}_{\mathrm{inter}}$ and $\widehat{\F}_{\mathrm{poly}}$ generally outperform $\widehat{\F}_0$ in feature augmentation, particularly for the two penalized linear models Lasso and Ridge Regression. However, different factors may stand out in different algorithms and for different datasets. For instance, as shown in Section \ref{sec:chinese_result}, $\F_{\mathrm{fnn}}$ outperforms all the other factors for Chinese-text learning under Lasso, Ridge, and RF while the polynomial kernel factors $\F_{\mathrm{rbf}}$ are the best under GBT and FNN. Therefore, based on the data and algorithms, it is recommended to try different factors to achieve the best performance.


\end{itemize}

\paragraph*{PCA and Diversified Projection}
Though only using a tiny subset of data ($1\%-5\%$ of the training samples) to conduct factor estimation, the diversified projection produces results comparable to those obtained through PCA on the entire training set. 
This can be seen in Figure \ref{Fig_show} where most black `+' points (for diversified projection) are at similar vertical positions of the corresponding colorful bars (for PCA).
With the efficiency of diversified projection guaranteed, it is possible to handle cases where the sample size or dimension is ultra-large. The Chinese news text data serves as an example of the effectiveness of diversified projection on datasets with an extremely large number of samples (around $10^6$). 
As shown by Figure \ref{Fig_Chinese_new}, it is obvious that under all the studied models, there exists two to five augmentation method achieved by diversified projection that significantly improves prediction accuracy.

Although diversified projection performs comparably to applying PCA on the entire training set, there are instances where the estimation is less accurate. One such example can be seen in the lower-left corner of Figure \ref{Fig_show}, where diversified projection is used on the MNIST dataset with an FNN model. The reason is that diversified projection only requires the diversified weight matrix $\W$ to have angles with the loading matrix $\mathbf{B}$ so that $\mathbf B$ is not diversified away. This is a trade-off between accuracy and computational efficiency, and by increasing the parameter $n^\prime$, the performance of diversified projection can be improved, as this brings it closer to PCA.

\begin{figure*}[htb]
	\begin{center}
		\captionsetup{width=\linewidth}
		\centerline{
			\subfigure{\includegraphics[width=0.48\linewidth]{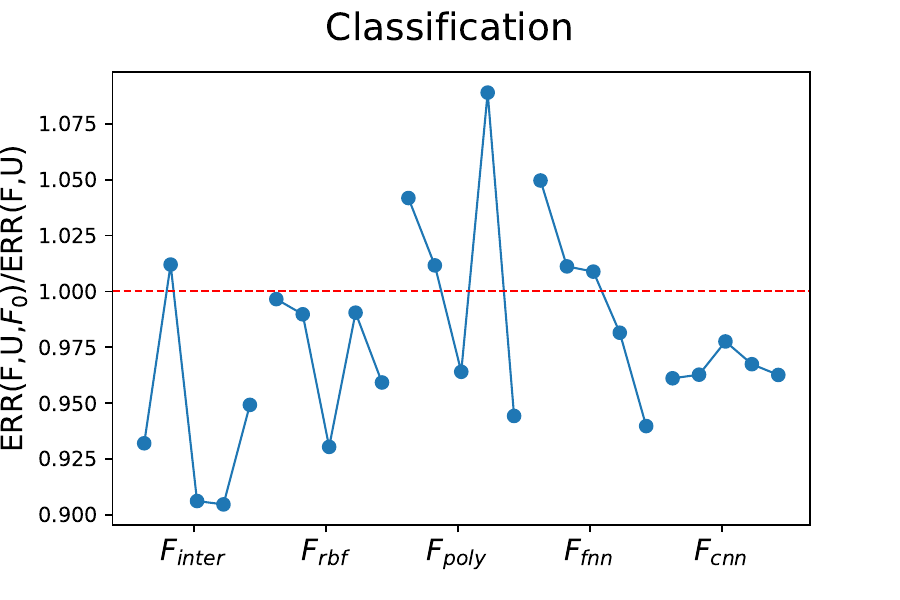}}
			\subfigure{\includegraphics[width=0.48\linewidth]{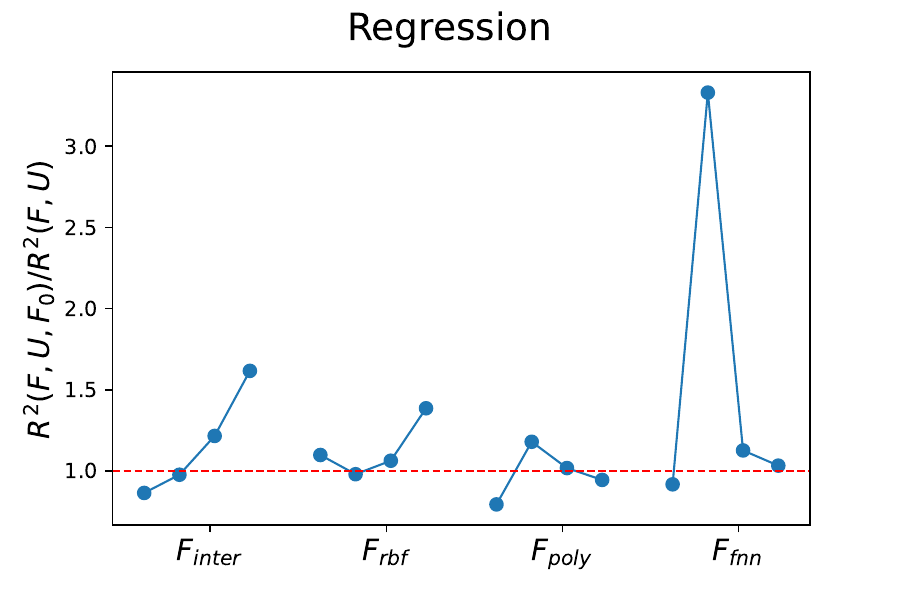}}
		}
		\caption{Comparison between PCA performances of $(\widehat{\F}_0,\widehat{\F},\widetilde{\U})$ and $(\widehat{\F},\widehat{\U})$. Left: Ratio of the classification error (ERR) of $(\widehat{\F}_0,\widehat{\F},\widetilde{\U})$ to the ERR of its corresponding $(\widehat{\F},\widehat{\U})$, averaging over all the classification problems. For each factor group, from left to right are results for Lasso, Ridge, RF, GBT, and FNN.
			Right: Ratio of the out-of-sample $R^2$ of $(\widehat{\F}_0,\widehat{\F},\widetilde{\U})$ to that of its corresponding $(\widehat{\F},\widehat{\U})$, averaging over all the regression problems. For each factor group, from left to right are results for Lasso, Ridge, RF, and GBT.
		}\label{Fig_QF}
	\end{center}
\end{figure*}

\paragraph*{Feature Augmentation Aggregation}
As has been mentioned in Section 2.3, different augmentation features may carry different information, so they may cooperate to enhance estimation performance. To see this, we first examine if different factors can cooperate to further boost the performance by adding $\widehat{\F}_0$ to the feature space $(\widehat{\F},\U)$. Figure \ref{Fig_QF} compares the performance of factor augmentations with and without $\F_0$. On the left-hand side, for each factor model and algorithm, we take the average over all the studied classification problems. If the point is below the one horizon line, it means that the factor contains different information from $\F_0$, and that $\F_0$ and the factor can cooperate to enhance the classification performance.  
Similarly, on the right-hand side, we look at the out-of-sample $\R^2$ averaging over all the studied regression problems for each factor model and algorithm. If the point is above the one horizon line, it means that $\F_0$ can cooperate with the corresponding factor to boost the regression performance.
It is presented that further augmenting with $\widehat{\F}_0$ can most of the time result in a little performance improvement, with an average reduction in classification error and an average increase in regression out-of-sample $R^2$.
Besides adding $\F_0$, it is also possible to use alternative or additional factors to aggregate information and enhance performance.

\begin{figure*}[htp]
	\begin{center}
		\captionsetup{width=\linewidth}
		\centerline{
			\subfigure{\includegraphics[height=4.2cm]{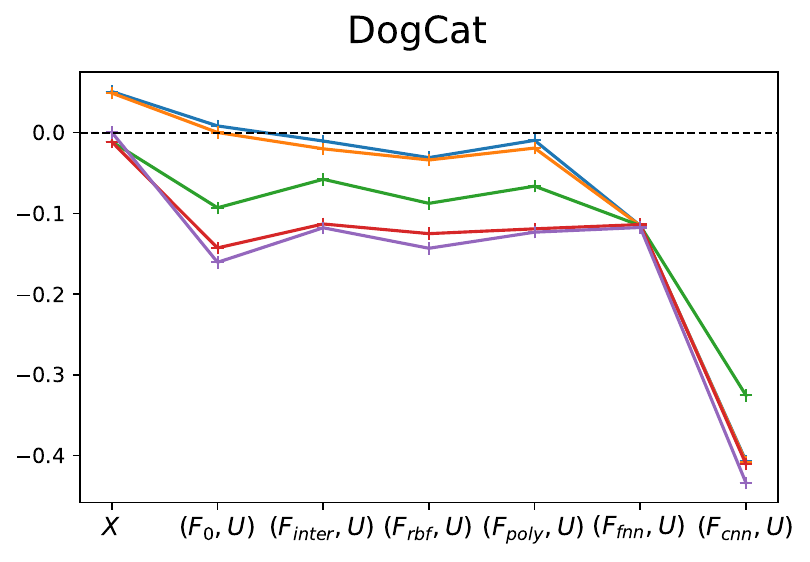}}
			\hspace{0.2cm}
			\subfigure{\includegraphics[height=4.2cm]{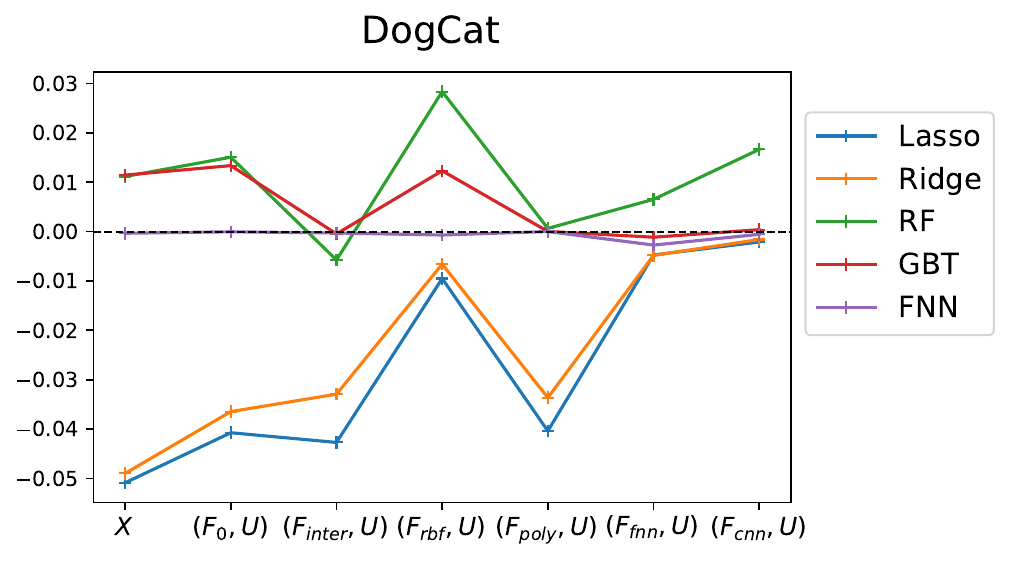}}
		}
		\caption{Left: Difference between the relative errors of factor augmentations and likelihood ratios for the DogCat dataset with PCA. For each algorithm and factors $(\F,\U)$, we present $(\operatorname{ERR}(\F,\U)-\operatorname{ERR}(\operatorname{LR},\X))/\operatorname{ERR}(\X)$.
			Right: Difference between the relative errors of different models with and without the likelihood ratios for the DogCat dataset with PCA. For each algorithm and factors $(\F,\U)$, we present $(\operatorname{ERR}(\operatorname{LR},\,(\F,\U))-\operatorname{ERR}(\F,\U))/\operatorname{ERR}(\X)$.}
		\label{Fig_LR2}
	\end{center}
\end{figure*}

Furthermore, likelihood ratios (LRs) are added as another feature augmentation technique to DogCat, the binary sentiment classification dataset. The performance of LR augmentation is compared to the proposed factor augmentations in the left-hand side of Figure \ref{Fig_LR2}. If the point in the plot is lower than the zero horizon line, it indicates that the factor augmentation outperforms the LR augmentation outperform. The left-most parts of the plot show that LR augmentation improves the performance of $\X$ in Lasso and Ridge regression, but barely in the other three studied algorithms. The comparison between LR augmentation and factor augmentation reveals that factor augmentations outperform LR augmentations on the DogCat dataset.
Moreover, the right-hand side of Figure \ref{Fig_LR2} displays the relative classification errors of different factor augmenting models with and without LR. If the point is below the zero horizon line, it indicates that further adding LR improves the performance in addition to the corresponding factor augmentation. It is shown that LR can further boost feature augmentation performance when using Lasso and Ridge, the two simple methods.
Nevertheless, it should be noted that while LR augmentation adds the same number of features as the original dataset, the proposed factor augmentations only add a limited number of features and can be applied to various problems, not just binary sentiment classification. Therefore, based on the results, factor augmentations can achieve similar or even better performance in some cases compared to LR augmentations with way fewer additional features and a broader scope of application.

\paragraph*{Cases Where Factor Augmentation Does Not Work Well}
For most cases, feature augmentations do boost the estimation performance, as shown in Figure \ref{Fig_show}. Nevertheless, there exist cases where factor augmentation does not work well. For the classification problems, when the sample size is too small, such as the Cervical cancer example (see Figure \ref{Fig_image}), which only has 72 samples in total, some feature augmentation approaches may not be superior compared to estimating directly with $\X$ under the advanced learning methods RF, GBT, and FNN. 
Since these three learning methods are much more complicated than Lasso and Ridge, it makes sense that adding some features does not boost the performance under those methods, especially when the sample size is small. However, even for the Cervical cancer dataset, we notice that the proposed feature augmentation approaches beat $\X$ under Lasso and Ridge. Meanwhile, with RF and GBT, there are always some different factors that defeat $\X$.  However, in the case of FNN, none of the factor augmentations outperformed $\X$. This is likely due to the fact that neural networks are capable of learning features themselves, and for the Cervical cancer dataset, the neural networks have already learned sufficient information even with only two hidden layers, rendering feature augmentation unnecessary.

\begin{figure*}[!htb]
	\captionsetup{width=\linewidth}
	\subfigure{\includegraphics[width=1.1\textwidth]{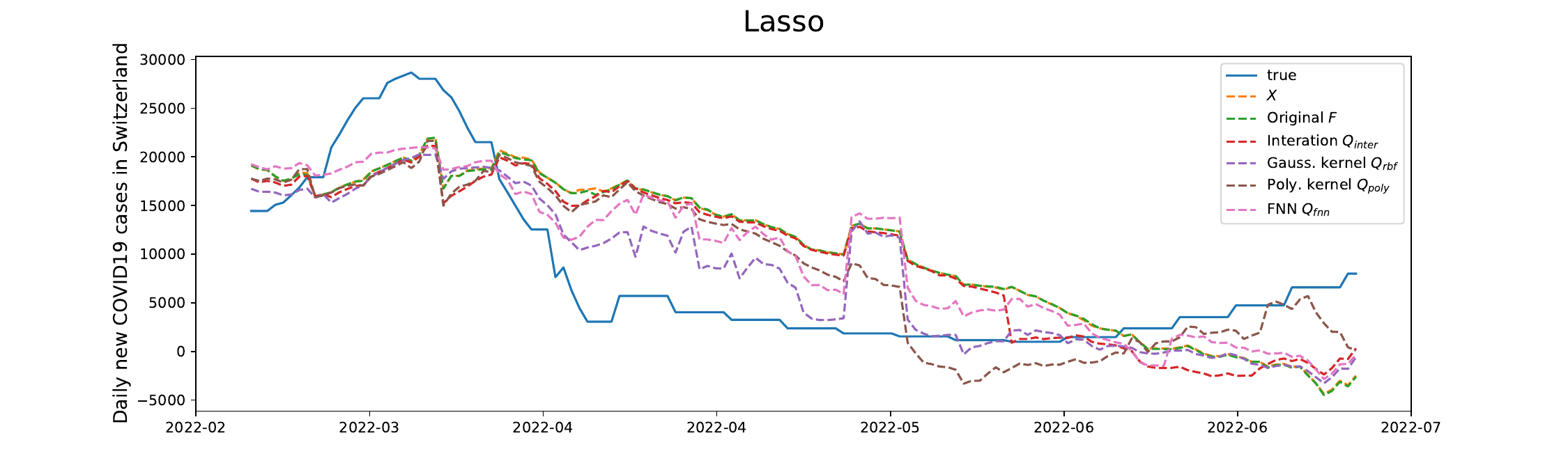}}
	\subfigure{\includegraphics[width=1.1\textwidth]{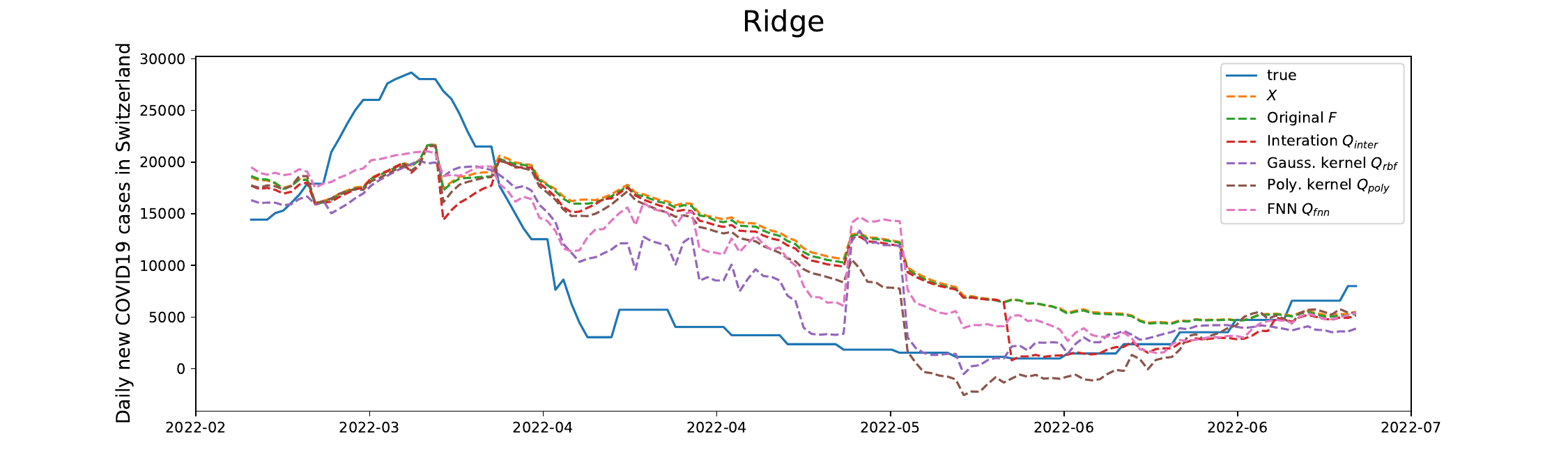}}
	\subfigure{\includegraphics[width=1.1\textwidth]{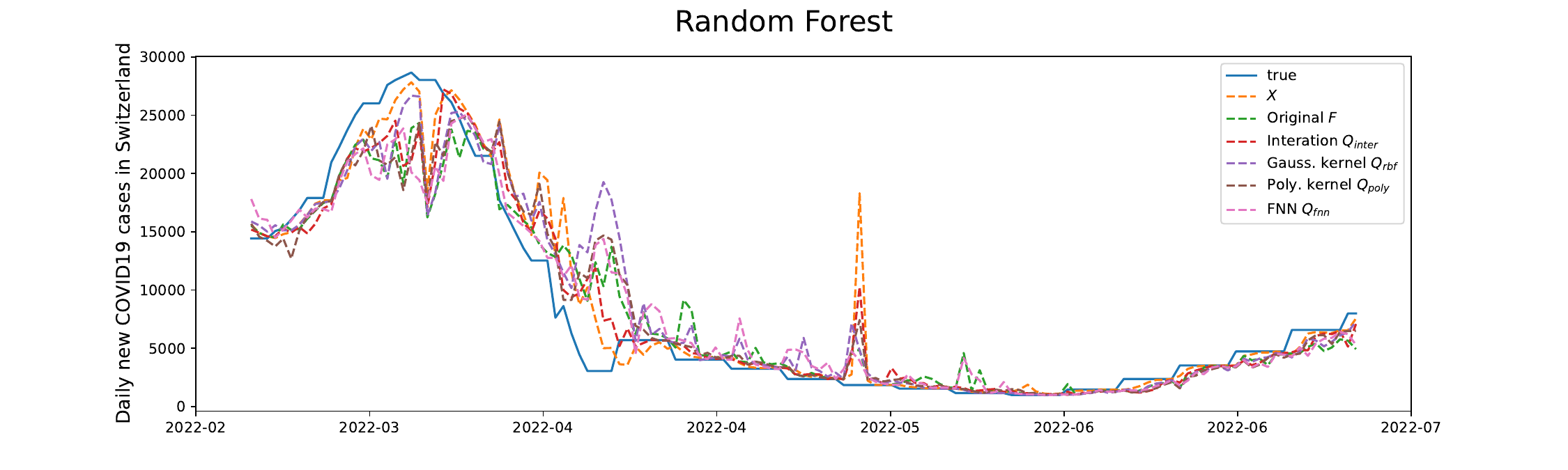}}
	\subfigure{\includegraphics[width=1.1\textwidth]{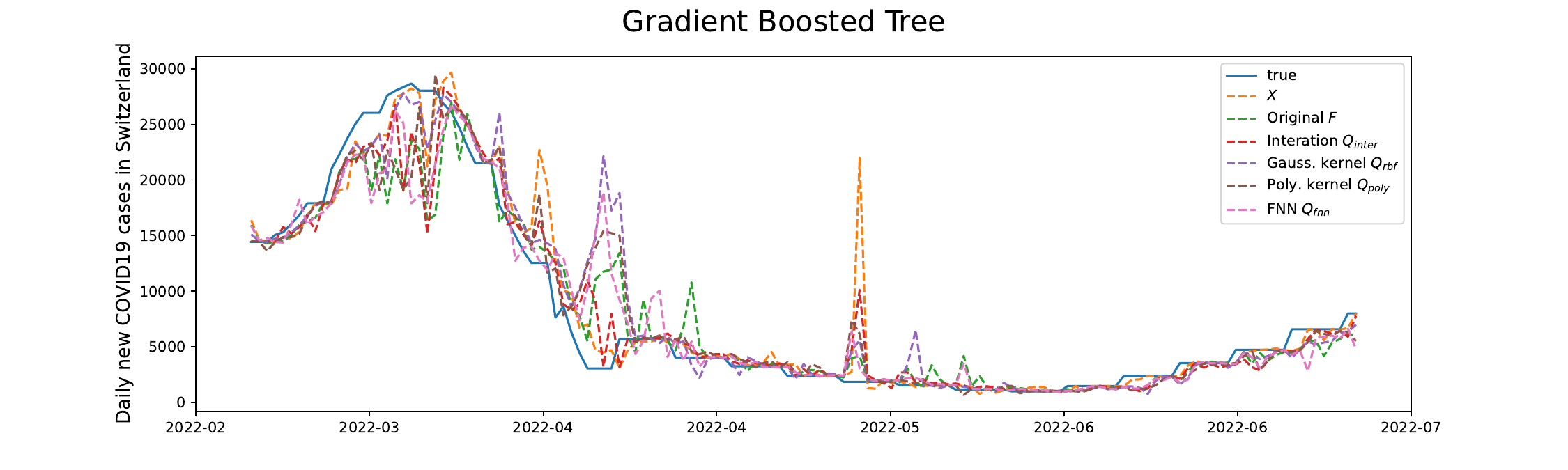}}
	\caption{One day ahead rolling window forecast for the daily COVID-19 new cases in Switzerland under the four algorithms. The dashed lines present the estimation based on $\X$, $(\widehat{\F}_0,\widehat{\U})$, and $(\widehat{\F},\widehat{\U})$ for the four different factors $\widehat{\F}$}\label{Fig_covid}
\end{figure*}

In some regression problems, certain algorithms may also fail to benefit from feature augmentations. For bond risk premia prediction, while most factor models can significantly boost performance in Lasso, RF, and GBT, Ridge regression does not perform well here because the feature dimension is of similar order with sample size, while Ridge does not induce sparsity. Besides, in cases where factor augmentations do not significantly improve performance, such as the Covid-19 data with RF and GBT, the taxi data, and the Zillow data, it can be seen that the regression on $\X$ already achieves satisfactory accuracy. For example, using RF or GBT to predict new Covid-19 cases with $\X$ already achieves an out-of-sample $R^2$ ranging from $0.93$ to $0.99$. This is accurate enough, and therefore, additional augmentation techniques are not necessary. (A visualization of the predictions on Covid19 in Switzerland is shown in Figure \ref{Fig_covid}.) What we have shown is that when the estimation is not that satisfactory, factor augmentations can bring about notable improvement and increase accuracy.

\paragraph*{Selection of transformation methods} To make the proposed framework more accessible to practitioners, we provide practical guidance on selecting appropriate nonlinear transformations. Exhaustively trying all available options is neither computationally efficient nor practically feasible. Instead, the following aspects should be considered when choosing suitable transformations:

\begin{itemize} 
\item \textbf{Characteristics of the problem and data:} The choice of transformation should align with the nature of the dataset and the domain-specific characteristics of the task. For instance, neural networks are known to be effective for textual data, while convolutional neural networks (CNNs) excel with image-based inputs. This intuition is consistent with our empirical findings, where $\F_{\mathrm{fnn}}$ performs well on the Chinese text data, and $\F_{\mathrm{cnn}}$ performs well on image datasets. Beyond such domain-specific preferences, more general considerations also apply: if interactions between samples are believed to be informative, kernel-based methods may be appropriate; if interactions among features are potentially meaningful, then constructing feature interaction matrices may be beneficial.

\item \textbf{Dimensionality of the original design matrix:} Although the proposed framework is already computationally efficient, extremely high-dimensional data may still necessitate caution. When the sample size is very large, full kernel matrices may be infeasible to compute, in which case other transformations or sparse kernel representations, such as the one used for the Chinese text data, are recommended. Conversely, when the number of features is very large, computing all pairwise interactions may become intractable, and techniques such as sure screening (see Section~\ref{sec:screen}) or alternative transformations should be considered.

\item \textbf{Complexity of the machine learning algorithm:} As shown in our experiments, the effectiveness of a transformation may depend on the complexity of the base learning algorithm. For example, in the Chinese text data application, $\F_{\mathrm{fnn}}$ substantially outperforms other transformations when used with simpler models such as Lasso, Ridge, or RF. However, this advantage diminishes under more complex models like GBT and FNN. In general, when using a complex learning algorithm, simpler transformations may suffice, while more sophisticated transformations are more useful when the learning model is relatively simple, such as linear regression.

\item \textbf{Need for interpretability:} While complex transformations such as those derived from neural networks (e.g., $\F_{\mathrm{fnn}}$) may offer strong predictive performance, they often lack interpretability due to their black-box nature. In domains where model transparency is critical — such as finance or econometrics — transformations that yield more interpretable factors may be preferable.
\end{itemize}

Although there is no universally optimal transformation method for a given dataset or algorithm, the considerations above provide a structured framework for narrowing the search space. Once a shortlist of plausible transformations is identified, the final selection can be made based on validation performance, akin to standard model selection procedures. Lastly, we note that it is also possible to combine multiple sets of factors or augmented features within a single model, as discussed at the beginning of this Section.

\paragraph*{Highlights} Detailed as has been illustrated above, we highlight three takeaway messages as follows. 
First, as an additional and independent method, our proposed augmentation strategy is designed not to compete but to collaborate with existing algorithms and possibly other data augmentation methods.
Second, it is evident that feature augmentation by various types of factors usually leads to improved performance, and the degree of improvement depends on the context of the problems. Not all feature augmentation methods will bring significant improvements, but there are always some factor augmentations that perform reasonably well.
Finally, the proposed factor augmentation approaches usually tend to be powerful when the initial estimation does not have high accuracy and the sample size is not exceedingly small.  This is understandable as they are fundamental limits on generation errors.

\section{Conclusion And Futher Discussion}\label{sec_conclusion}
We have put forward a series of simple yet effective feature augmentation techniques to study the stock return prediction using Chinese financial news text data, with a generalization to all high-dimensional learning problems. 
These methods are based on extracting latent nonlinear factors through transformations of the original feature matrix. We illustrate three representative transformation families — interactions, kernel methods, and neural networks — due to their potential to uncover meaningful latent structures in text data in the stock return prediction problem. While these serve as core examples, the framework is flexible and can accommodate a wide range of transformation techniques, provided they are capable of capturing useful structure informed by the nature of the data.

Our empirical analysis demonstrates that augmenting the feature space with nonlinear factors significantly enhances the accuracy of stock return estimation, which in turn leads to consistent improvements in downstream financial applications, including event studies and portfolio construction. To further validate the generality of the approach, we conduct extensive experiments across diverse supervised learning problems, including both classification and regression tasks from domains such as image recognition, biology, finance, and natural language processing. These results affirm the versatility and practical utility of the proposed framework. We offer a broad set of factor types and augmentation strategies, and provide practical guidance for selecting suitable transformations based on problem-specific considerations. In the majority of studied cases, factor-based augmentation yields notable gains in generalization performance with minimal additional computational cost.

A key strength of the proposed framework lies in its modularity: it operates independently of the underlying learning algorithm and can be seamlessly integrated into any method that leverages covariate information. Moreover, the augmentation need not be confined to the initial input layer—it can also be applied at intermediate stages of learning pipelines, particularly in settings with highly correlated features or weak marginal signals. Also, while our implementation primarily relies on linear factor extraction techniques, such as principal component analysis and diversified projections, the framework is equally compatible with nonlinear alternatives, like autoencoder-decoders \citep{xiu2024deep, cerqueti2024improving}. For instance, one may construct an hourglass-shaped FNN, e.g., with hidden layer widths $[128, 64, 16, 64, 128]$, and train it to reconstruct the transformed design matrix. The bottleneck layer, which compresses the high-dimensional input into a low-dimensional representation, can then be treated as a set of learned nonlinear factors. Such approaches offer a promising direction for future research, particularly in capturing complex structures beyond the reach of linear models.

Finally, this augmentation strategy readily extends to matrix- and tensor-valued data, supported by recent developments in high-dimensional factor modeling \citep{chen2024factor, chen2022factor}. These extensions enhance the relevance and applicability of our framework in emerging areas such as macroeconomic forecasting, financial modeling, and biological data analysis.

\clearpage

\clearpage
\bibliographystyle{imsart-nameyear} 
\bibliography{main}       

\end{document}